\documentclass[twocolumn]{aastex631} 

\usepackage{graphicx}
\usepackage{graphbox}
\usepackage{hyperref}
\usepackage{amssymb,amsmath}
\usepackage{stmaryrd}
\usepackage{booktabs}
\usepackage{gensymb}

\usepackage[print-unity-mantissa=false, separate-uncertainty=true, multi-part-units=single]{siunitx}
\DeclareSIUnit\erg{erg}

\def\lea{\mathrel{<\kern-1.0em\lower0.9ex\hbox{$\sim$}}}
\def\gea{\mathrel{>\kern-1.0em\lower0.9ex\hbox{$\sim$}}}
\newcommand{\lta}{{\>\rlap{\raise2pt\hbox{$<$}}\lower3pt\hbox{$\sim$}\>}}
\newcommand{\gta}{{\>\rlap{\raise2pt\hbox{$>$}}\lower3pt\hbox{$\sim$}\>}}

\usepackage{xcolor}

\usepackage{amsmath} 
\usepackage{floatrow}

\begin{document}

\title{The nature of UHE source 1LHAASO J1740+0948u and its connection to PSR J1740+1000}

\shorttitle{The nature of UHE source 1LHAASO J1740+0948u}

\author[0000-0003-0902-1935]{Seth Gagnon}
\affiliation{Department of Physics, The George Washington University, 725 21st St, NW, Washington, DC 20052, USA}

\author[0009-0009-1950-6937]{Yichao Lin}
\affiliation{Department of Physics, The George Washington University, 725 21st St, NW, Washington, DC 20052, USA}

\author[0000-0003-3540-2870]{Alexander Lange}
\affiliation{Department of Physics, The George Washington University, 725 21st St, NW, Washington, DC 20052, USA}

\author[0000-0002-8832-6077]{Hui Yang}
\affiliation{Institute for Research in Astrophysics and Planetology (IRAP), CNRS, Toulouse, 31400, France}

\author[0000-0002-7465-0941]{Noel Klingler}
\affiliation{Center for Space Sciences and Technology, University of Maryland, Baltimore County, Baltimore, MD, 21250, USA}
\affiliation{Astrophysics Science Division, NASA Goddard Space Flight Center, Greenbelt, MD, 20771, USA}
\affiliation{Center for Research and Exploration in Space Science and Technology, NASA Goddard Space Flight Center, Greenbelt, MD, 20771, USA}

\author[0000-0002-8548-482X]{Jeremy Hare}
\affiliation{Astrophysics Science Division, NASA Goddard Space Flight Center, Greenbelt, MD, 20771, USA}
\affiliation{Center for Research and Exploration in Space Science and Technology, NASA/GSFC, Greenbelt, Maryland 20771, USA}
\affiliation{The Catholic University of America, 620 Michigan Ave., N.E. Washington, DC 20064, USA}

\author[0000-0002-6447-4251]{Oleg Kargaltsev}
\affiliation{Department of Physics, The George Washington University, 725 21st St, NW, Washington, DC 20052, USA}

\begin{abstract}

We present multi-wavelength 
analysis of 1LHAASO J1740+0948u 
and its surroundings including the
pulsar wind nebula of middle-aged pulsar PSR J1740+1000. Although 
a dozen X-ray sources are found within the UHE emission site,  careful analysis shows that they are unlikely to produce the observed UHE emission.  
The most likely particle  accelerator 
is
pulsar J1740+1000
which if offset by 13$'$ 
north of the UHE source but 
appears to be 
connected to it by an extended  feature seen in X-rays.  
For a plausible pulsar distance of 1.2 kpc, 1LHAASO J1740+0948u must be  located about 5 pc away which requires rapid transport of electrons along the feature to avoid radiative losses.
This poses several challenges for standard pulsar theory. Firstly, being produced $\lesssim10$ kyrs ago,  
particles must have been accelerated to the energy corresponding to 
a large fraction of the pulsar's full potential drop across the polar cap. Secondly, due to the lack of TeV emission extension toward the pulsar, particles must
be accumulating in the UHE region. In this context, we discuss two possible scenarios: a tail filled with pulsar wind and confined by the bow-shock due to the fast pulsar's motion and 
an ISM filament filled by the most energetic pulsar wind 
particles escaping from the apex of the bow-shock.

\end{abstract}

\section{\label{sec:intro}Introduction}

Pulsars produce powerful winds of energetic particles that interact with their environment.
Downstream of the termination shock the ultra-relativistic pulsar wind 
flow slows down abruptly and particle radiation 
becomes observable across the electromagnetic (EM) spectrum  as a Pulsar Wind Nebula (PWN). 
At larger distances, the post-shock flow  is expected to transition from bulk flow (advection) to 
diffusive 
particle transport
and can be greatly affected by its interaction with either the supernova remnant (SNR) reverse shock, or by the ram pressure of the oncoming medium if the pulsar is moving very fast. For older pulsars that escaped from their host SNRs, highly supersonic motion is expected, leading to the formation of a bow-shock and a pulsar tail (see \citealt{2017SSRv..207..175R,2023Univ....9..402O} for reviews).    

In PWNe ultra-relativistic particles 
gyrate along magnetic field lines,
producing synchrotron emission
from radio energies all the way up to GeV 
(see e.g., \citealt{2015SSRv..191..391K}).
At even higher energies ($>1$ TeV)
photons can be produced by  
leptonic and hadronic processes (see e.g., \citealt{2021Natur.594...33C, 2024arXiv240210912A}).
In the leptonic scenario, relativistic electrons in the pulsar wind 
interact with the ambient photon fields, such as the Cosmic Microwave Background (CMB), upscattering them via Inverse Compton (IC)
to TeV energies.
In this scenario, these same electrons will be producing 
synchrotron emission at lower energies.
In the hadronic scenario, 
relativistic protons interact with the dense environment 
producing $\pi^0$ 
particles that then decay, producing TeV gamma-rays via
$\pi^0 \rightarrow \gamma + \gamma$.
In this scenario, the synchrotron emission could be produced by the secondary $e^{+}/e^{-}$ pairs (from the accompanying charged pion decay process) and can be fainter (than in the leptonic scenario).
Although, 
a small fraction of relativistic protons (ions)  
could,  
in principle, be present in the pulsar wind, 
 the hadronic scenario is more often discussed in the context of a SNR shock interacting with nearby dense clouds. 
For this reason, the leptonic mechanism is the relevant one to consider for older pulsars moving in a low-density ISM.
Within the leptonic scenario, PWN emission in ultra-high energy (UHE) $\gamma$-rays ($>100$ TeV) 
can be used to directly probe the maximum energy to which $e^{+}/e^{-}$ pairs are accelerated in the pulsar wind \citep{2022ApJ...930L...2D}.
Thus, UHE observations 
provide a window into the poorly understood acceleration mechanism operating in pulsar winds as well as ultra-relativistic particle  transport 
in the ISM (see \citealt{2024arXiv240210912A} for recent review).

The Large High Altitude Air Shower Observatory (LHAASO) 
is a part of the new generation of ground-based 
arrays
with greatly improved sensitivity,
allowing for robust source detection up to 1.4 PeV.
The LHAASO collaboration
recently published a catalog 
which includes 43 
UHE sources detected above $>$100 TeV \citep{2024ApJS..271...25C}.
One of these 
UHE sources,
LHAASO J1740+0948u 
($l = 33.800\degree, b = +20.260\degree$), is located $\sim13'$ southwest of a 
100-kyr-old radio pulsar
(discovered by \cite{2002ApJ...564..333M})
PSR J1740+1000 (hereafter J1740) and  has several X-ray sources within its $r\approx 4'$ statistical positional uncertainty (PU, at the 95\% confidence level),
but no confidently established counterpart thus far.
The pulsar's 
period and period derivative ($P=154.1$ ms,  $\dot{P}=2.15\times10^{-14}$ s s$^{-1}$) 
suggest that it is an unremarkable middle-aged  pulsar 
with a  
spin-down energy loss rate $\dot{E}=2.3\times10^{35}$ erg s$^{-1}$ and surface magnetic field $B_{0}=1.8\times10^{12}$ G. Pulsed GeV emission with 154-ms period was recently detected with LAT detector onboard Fermi Gamma-ray Observatory \citep{2022MNRAS.513.3113R, 2023ApJ...958..191S}.   
At the dispersion measure (DM)
distance\footnote{According to the ATNF pulsar catalog \citep{2005AJ....129.1993M}, both models of electron density distribution \citep{2002astro.ph..7156C,2017ApJ...835...29Y}, give similar distances of 1.22 and 1.24 kpc. } of $\simeq1.2$ kpc
and Galactic latitude of
$20.3^\circ$, J1740
is located well above the Galactic
plane. 
This  makes the hadronic scenario less likely due to the decreased ISM density at this latitude and lack of dense clouds 
in the region \citep{2025arXiv250215447C}. In addition,   
 the modeling of the IC component  from the associated PWN is simplified because the relevant (for TeV energies) external radiation field energy density is dominated by the CMB. Therefore, if the recently discovered UHE source LHAASO J1740+0948u  \citep{2025arXiv250215447C} is  powered by the J1740 pulsar, it could be an excellent testbed for models of evolved PWNe \citep{2023Univ....9..402O} given that most other UHE LHAASO sources are located within more complex environments in the Galactic plane.  The maximum energy of TeV photons detected from LHAASO J1740+0948u is  $\sim300$ TeV. The TeV source is also reported in the 3rd HAWC observatory catalog \citep{2020ApJ...905...76A} as 3HWC J1739+099 with position consistent with the LHAASO source position. However, TeV emission was not detected by VERITAS 
 with the upper limits
 \citep{2021ApJ...916..117B}
 on flux close to HAWC flux measurements  
 \citep{2020ApJ...905...76A}.
 The LHAASO's spectrum of the source can be described by a power-law (PL) with photon index $\Gamma\approx3$ but a slightly better fit can be obtained with log-parabola model indicating possible curvature. \cite{2025arXiv250215447C} do not explore a fit using an exponentially cutoff PL model which could, potentially, also account for the curvature. 
 The compactness of the UHE source (the source is unresolved by LHAASO with its size limited to $<9'$) 
 and its large distance from the pulsar 
 argue against the purely isotropic diffusion model often invoked to explain the so-called TeV halos around older pulsars \citep{2024PhRvD.109h3026D}.
 More appropriate scenarios would be those where particles are either advected along the pulsar tail (i.e.\ organized bulk flow; \citealt{2021ApJ...916..117B})  or stream along the ISM magnetic field directed from the pulsar to the UHE source (the so-called ``misaligned outflow'' (MAO) or ``pulsar filament'' scenario; \citealt{2022ApJ...928...39D,2023ApJ...950..177K}). The latter two scenarios are supported by the existence of the elongated diffuse X-ray feature, discovered by \cite{2008ApJ...684..542K},  
extending from the pulsar toward the UHE source.
 In either of these scenarios, the TeV emission can be brighter farther away (along the tail) from the pulsar if particles accumulate downstream as the flow slows (cf.\ AGN lobes; e.g., \citealt{Croston2005,Hardcastle2020}). Similar spatial offsets between synchrotron and inverse-Compton components have also been noted in other systems such as microquasars (e.g., \citealt{Bordas2009}) and the recently identified ``Peanut'' structure (\citealt{2025arXiv251006786C}).
 An alternative explanation for the offset is that LHAASO J1740+0948u 
 may not be related to the J1740 pulsar and then a different counterpart can be found at lower energies. 
 
 In this paper we explore both scenarios. In Sections \ref{sec:xray_obs} and \ref{sec:Fermi_analysis}, we describe X-ray and gamma-ray observations taken with XMM-Newton and Fermi-LAT and their respective data
 analyses.
 We discuss their implications for LHAASO J1740+0948u  in Section \ref{sec:discussion} and conclude with a summary of the findings in Section \ref{sec:summary}. 

\section{\label{sec:xray_obs}X-ray observations}

J1740 was observed with {\sl XMM-Newton} 
on multiple occasions 
between 2017 September 20 and 2018 April 04
(obsID's: 0803080201
0803080301
0803080401
0803080501),
with a total exposure of 532 ks. 
However, the  LHAASO source region was  partly covered by only  MOS2 (due to EPIC pn being operated in Small Window mode and MOS1 having a  dead chip) while also  placed far off-axis.
The combined image of the pulsar's vicinity (imaged with 
both MOS detectors)
revealed an extended feature
clearly visible up to $\approx6'$ SW of the pulsar.

The {\sl XMM-Newton} data were
processed using the standard {\tt emproc} routine from the XMM
Science Analysis Software (SAS; version 21.0). We manually
filtered the data for periods of high background and flaring.
Spectra from each observation and corresponding response
files were extracted and produced using the standard routines, 
{\tt evselect}, {\tt arfgen}, and {\tt rmfgen}, and grouped to 50
counts per bin using {\tt specgroup}.
For each observation
the spectra 
extracted from the MOS1 and MOS2 detectors 
were fit
simultaneously
using XSPEC (\citealt{1996ASPC..101...17A}, version
12.13.0c). In all spectral fits we
restricted the energy range to 0.5 -- 10 keV
and
used a simple absorbed
power-law (PL) model (XSPEC’s {\tt tbabs$\times$pow}), 
with 
solar abundances from
\citealt{2000ApJ...542..914W}
and photoionization cross-sections from \citealt{1996ApJ...465..487V}.
Unabsorbed flux was measured by freezing the absorbed PL parameters and adding the {\tt cflux} component.
All
uncertainties listed correspond to 1$\sigma$.

\begin{figure*}[t!]
\vspace{-0.0cm}
\centering
\hspace{-0.0cm}
\includegraphics[height=10.0cm,angle=0]{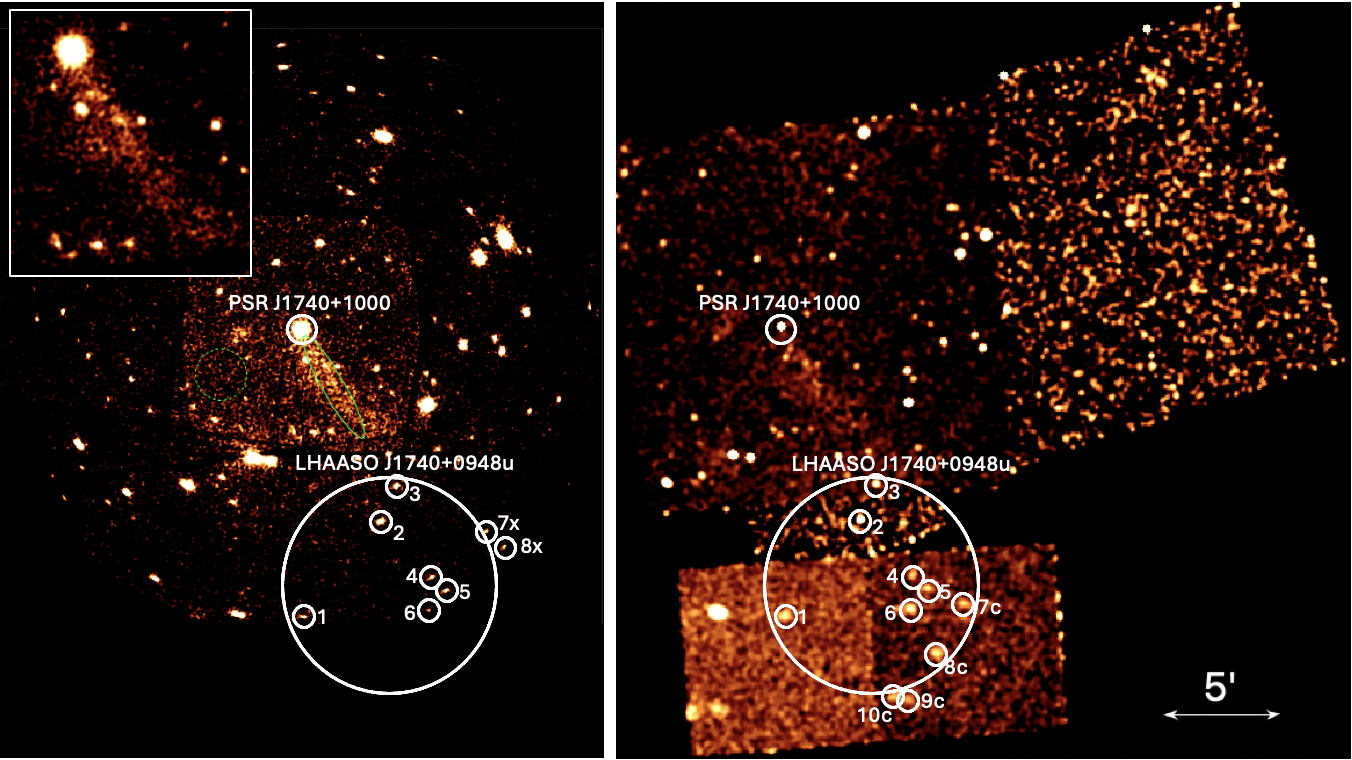}
\caption{\footnotesize
 X-ray images of  LHAASO J1740+0948u field. The image on the left is 
 from
 XMM-Newton EPIC MOS1+2 
 (0.5 -- 10 keV)
 and on the right is from CXO ACIS 
 (0.5 -- 7 keV).
 X-ray sources in the LHAASO field are labeled 
 in each image, and correspond to entries in Table \ref{tab:combine_table}. The regions used for spectral extraction of the pulsar tail
 in the XMM data
 are shown in green with solid (source) and dashed (background) lines.
 The inset on the left is a
 zoomed-in portion of the XMM image
 showing the pulsar and its tail.  
 }
\label{fig:xray_images}
\end{figure*}

The pulsar was also observed for 69 ks with Chandra X-ray Observatory (CXO) 
(obsID's: 1989, 11250).
In the ACIS image the tail is 
detected 
up to $\sim 3'$ from the pulsar. 
Although ACIS covered the part of the UHE source uncertainty circle that was not covered by {\sl XMM-Newton} observations, the very far ($\gtrsim 16'$) off-axis placement severely degrades the angular  resolution and sensitivity. Nonetheless, due to the very low ACIS background, 
CXO detected several sources (albeit with rather uncertain positions, see Figure \ref{fig:xray_images}).
We do not use CXO ACIS data for the tail spectrum because XMM-Newton data are superior due to much longer exposure. 
We, however, use CXO data below  to analyze sources within the UHE emission site. 

\subsection{\label{sec:xray_fits}
MOS spectra of the pulsar tail. }

The combined MOS1/2 image reveals a
feature  extending southwest up to approximately $6'$ from
the pulsar position. The initial segment of the 
outflow appears
to be conical, with an opening angle of $\approx 20 \degree $. 
The
spectrum of the extended feature 
was extracted from the region shown in the left
panel of Figure \ref{fig:xray_images}.
The spectrum was fit with an absorbed power-law (PL) model
with the
absorbing hydrogen
column density
frozen
at 
 n$_\mathrm{H} = 1 \times 10^{21} \mathrm{cm}^{-2}$ adapted by \cite{2008ApJ...684..542K}. 
The best fit photon index was
$\Gamma = 1.62 \pm 0.05$
with a normalization of
$\mathcal{N} = (1.41 \pm 0.05)\times10^{-5}$
photon s$^{-1}$
cm$^{-2}$ keV$^{-1}$ (at 1 keV),
yielding
the  reduced chi-squared 
$\chi^2_\nu$= 
1.06
(for $\nu$ =
463
degrees of freedom).
The unabsorbed 0.5--10 keV flux
is
$F_X = (9.8 \pm 0.7) \times 10^{-14}$ 
erg cm$^{-2} \mathrm{s}^{-1}$ 
corresponding to an
X-ray luminosity of
$L_X = 1.7 \times 10^{31}$
$\mathrm{erg
s}^{-1}$
(at $d=1.2$\,kpc), and an X-ray efficiency
of
$\eta_X =L_X/\dot{E} \approx 10^{-4}$. 

\subsection{\label{sec:xray_sources}X-ray Sources detected within the LHAASO source region}

Although LHAASO J1740+0948u lies at the continuation of the extended structure originating from the pulsar, 
it may be that the TeV emission is produced by a different source such as a blazar or even another pulsar. For example, 1LHAASO J0359+5406, where two middle-aged pulsars, J0359+5414 \citep{2018MNRAS.476.2177Z} and B0355+54 \citep{2006ApJ...647.1300M}, are located only $5'$ apart and either (or both) 
could be the TeV source's
counterparts given 
the TeV source's 
positional uncertainty (PU) \citep{2024ApJS..271...25C}. However, only  the latter pulsar 
is accompanied by an extended X-ray tail \citep{2016ApJ...833..253K}.

\begin{figure*}[!]
\centering
\includegraphics[width=18.0cm,angle=0]{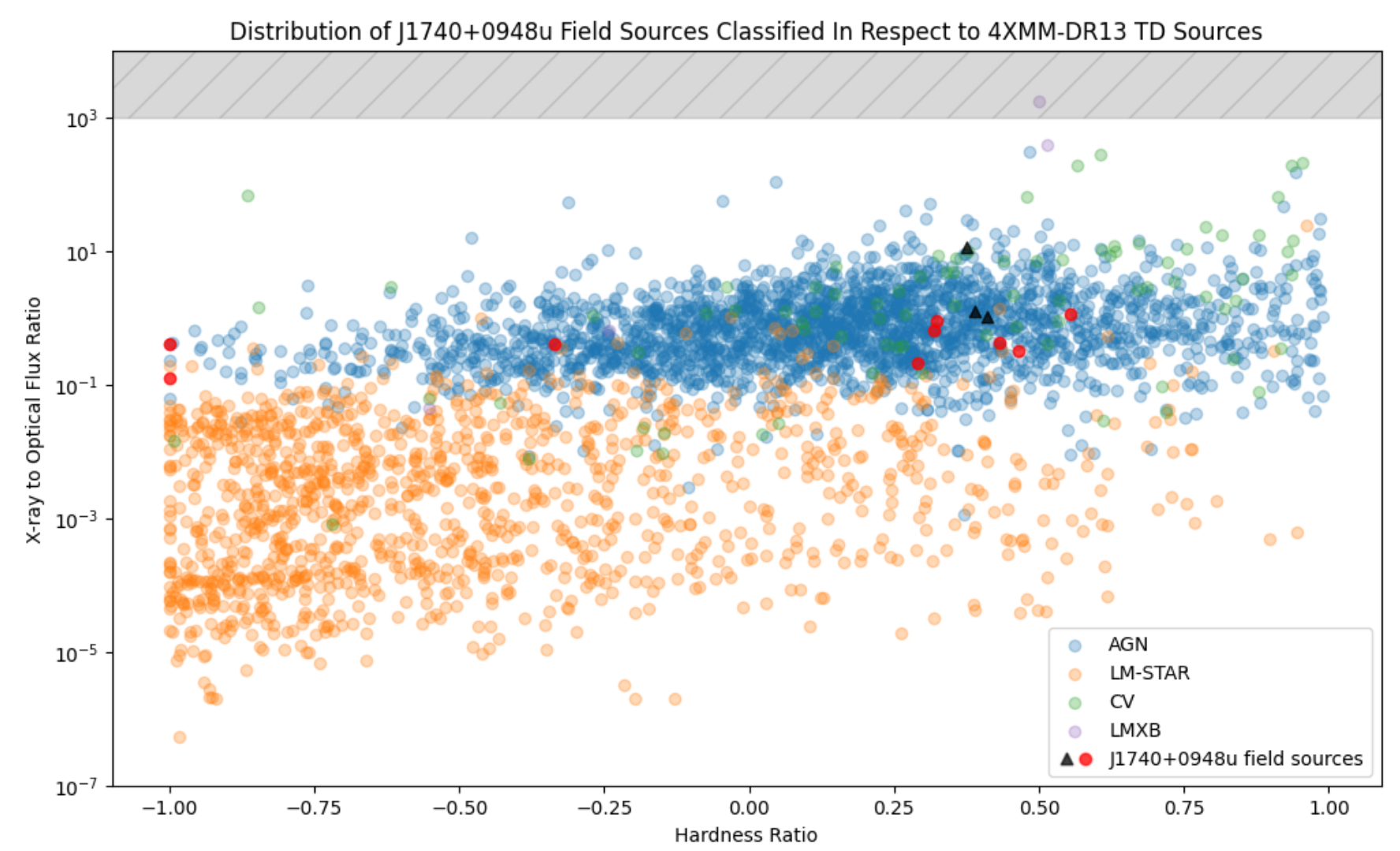}
\caption{\footnotesize
X-ray to optical flux ratios versus X-ray hardness ratios 
for 1LHAASO J1740+0948u field X-ray sources alongside 4XMM-DR13 training dataset X-ray sources \citep{2024RNAAS...8...74L}.
We performed cross-matching using {\sl XMM-Newton} positions to PanSTARRS and only retained X-ray sources with a single PanSTARRS match to avoid confusion.
Optical fluxes are obtained from the PanSTARRS g-band magnitude with the exception of PSO J264.9676+09.7932, the counterpart to 2CXO J173952.1+094733, where the r-band magnitude is used instead due to the lack of g-band magnitude. X-ray to optical flux ratios are  defined as $R_{\mathrm{x}/\mathrm{o}}=F_{0.2-12.0 \mathrm{keV}}/F_g$ for {\sl XMM-Newton} sources and $R_{\mathrm{x}/\mathrm{o}}=F_{0.5-7.0 \mathrm{keV}}/F_g$ for CXO sources. 
1LHAASO J1740+0948u field sources without PanSTARRS counterparts are shown as red circles and with black triangles representing the lower limits. The three sources with lower limits, src7x, src8x, and src8c have $R_{\mathrm{x}/\mathrm{o}}$ of 1.04, 1.24, 11.50, respectively. 
 The expected region for isolated pulsars is shaded in gray (based on 
 optically detected pulsars from \citealt{2013MNRAS.436..401M}).
Hardness ratio is defined as $\textrm{HR}=(F_{2.0-12.0 \textrm{ keV}}-F_{0.2-2.0 \textrm{ keV}})/(F_{2.0-12.0 \textrm{ keV}}+F_{0.2-2.0 \textrm{ keV}})$ for {\sl XMM-Newton} sources and $\textrm{HR}=(F_{2.0-7.0 \textrm{ keV}}-F_{0.5-2.0 \textrm{ keV}})/(F_{2.0-7.0 \textrm{ keV}}+F_{0.5-2.0 \textrm{ keV}})$ for CXO sources. 
For X-ray sources from the 1LHAASO J1740+0948u site, {\sl XMM-Newton} flux is used instead of CXO flux when available because the former is more accurately measured. 
}
\label{X_o_flux_HR_XMM}
\end{figure*}

To rule out the possibility of an unidentified X-ray source
being a counterpart to LHAASO J1740+0948u, we examined 
X-ray sources from the Chandra Source Catalog v2.1 (\citealt{2010ApJS..189...37E, 2020AAS...23515405E}, hereafter CSC v2.1) and 4XMM-DR13 Catalog \citep{2020A&A...641A.136W}
in the vicinity of LHAASO J1740+0948u
(see Figure \ref{fig:xray_images}). 
To start, we 
performed a multiwavelength (MW)  machine learning (ML) classification of these X-ray sources, using MUWCLASS, an ML  pipeline\footnote{See Appendix \ref{ML_appendix} for brief description of the classification pipeline.}
developed by    \cite{2022ApJ...941..104Y}. Prior to  classification 
these X-ray sources 
were cross-matched to Gaia DR3 \citep{2023A&A...674A...1G}, 
Two Micron All Sky Survey (\citealt{2006AJ....131.1163S}, hereafter 2MASS), ALLWISE \citep{2014yCat.2328....0C}, and
CatWISE \citep{2014yCat.2328....0C}.  The classification results 
are listed in Table \ref{tab:combine_table}.  We use the confidence threshold (CT) to 
characterize the reliability of classification, considering CT$>2$ to be confident classifications 
(see \citealt{2022ApJ...941..104Y} for details).
The MW counterparts of X-ray sources and their association probabilities 
can be found in Table A1 in Appendix \ref{ML_appendix}.
We also cross-matched the J1740+0948 field sources 
to PanSTARRS1 \citep{2016arXiv161205560C} and plotted the X-ray to optical flux ratios versus hardness ratios for the MUWCLASS training dataset (TD) sources and the 1LHAASO J1740+0948u field X-ray sources (see Figure \ref{X_o_flux_HR_XMM}). Notably, the 1LHAASO J1740+0948u field X-ray sources are located in the region occupied by active galactic nuclei (AGN). 
Out of 12 unique X-ray sources identified (see Table \ref{tab:combine_table}), 6 X-ray sources are 
detected with both CXO and {\sl XMM-Newton}, 4 are only 
detected by CXO, and 2 are only 
detected by {\sl XMM-Newton}. 
The non-detections by one or the other observatory are  
due to the lack of coverage (see Figure \ref{fig:xray_images}). 
All 7 confidently classified sources belong to the AGN class, in which only src1 has a disagreeing classification between CXO and XMM-Newton. This is likely due to the lack of MW counterparts for the CXO source, caused by the large ($4.95"$) offset between the CXO source position and the nearest MW counterpart, while the XMM source has a small ($0.24"$) offset to the nearest MW counterpart. The discrepancy between the  XMM-Newton and CXO positions is not surprising, given the extremely off-axis position of src1 in the CXO image, which hampers the accuracy of its localization and the reliability of its MW counterpart.
Only one source, 2CXO J174008.4+095259 (src3) is variable and is only variable in CSCv2.1. This source was observed in 2001 and 2010 with a per-observation broadband flux ({\tt o.flux\_aper90\_b}) of 
 2.14$\times10^{-14}\mathrm{ergs}\ \mathrm{cm}^{-2} \mathrm{s}^{-1}$ and 1.71$\times10^{-14}\mathrm{ergs}\ \mathrm{cm}^{-2} \mathrm{s}^{-1}$, respectively.
All of these sources lack significant proper motion (PM) measurements in Gaia DR3 and  CatWISE catalogs\footnote{If detected, the ratio of PM magnitude to  the measurement uncertainties $<$5)}. 
 One of these 7 sources (src4) is the only X-ray source with the  radio counterpart. 
 Three more sources,  
src2, src7x, and src8x, 
were classified as low-confidence AGN.  We note that two of these sources, src7x and src8x, are the only sources (out of 12)  lacking any MW associations. We discuss these 3 sources below in more detail.

\begin{table*}[t!]
\vspace{-0.1cm}
\caption{Classification of X-ray Sources in the J1740+0948u Field} 
\vspace{-0.5cm}
\setlength{\tabcolsep}{0.08in} \footnotesize{
\begin{center}
\begin{tabular}{lcccccccccccc}
\hline\hline & name & Class &  $P^a$ &  $e\_P^b$ & CT$^c$ &  \#\_CTP$^d$&      Sep$^e$ & Sig.$^f$ &  PU$^g$ &  F\_b$^h$ & HR$^i$ \\ 

 &   &   &    &    &   &   &     ($'$) &  & ($''$)  &  ($\mathrm{ergs}\ \mathrm{cm}^{-2} \mathrm{s}^{-1}$) \\ \hline

1 & 4XMM J174025.5+094703 &   AGN &    0.97 &      0.02 & 30.48 &        1 & 4.77 &  52.34 &   0.94 &  3.77$\times10^{-14}$ & 0.55 \\ 
2 & 4XMM J174011.2+095122 &   AGN &    0.39 &      0.22 &  0.09 &        0 & 2.95 & 166.81 &   0.46 &  3.07$\times10^{-14}$ & 0.29 \\ 
3 & 4XMM J174008.4+095258 &   AGN &    0.96 &      0.02 & 23.10 &        1 & 4.39 &  58.09 &   0.53 &   2.06$\times10^{-14}$ & 0.46 \\ 
4 & 4XMM J174001.8+094849 &   AGN &    0.97 &      0.03 & 16.36 &        1 & 1.33 &  90.47 &   0.52 &   2.36$\times10^{-14}$ & 0.32 \\ 
5 & 4XMM J173959.2+094814 &   AGN &    0.97 &      0.02 & 26.43 &        1 & 1.98 &  94.92 &   0.68 &   2.38$\times10^{-14}$ & 0.43 \\ 
6 & 4XMM J174002.3+094721 &   AGN &    0.99	&      0.02 & 33.16 &        1 & 1.73 &  74.10 &   0.61 &   3.53$\times10^{-14}$ & 0.32 \\ 
7x & 4XMM J173951.8+095055 &   AGN &    0.58 &      0.25 &  0.78 &        0 & 4.45 &  35.40 &   0.66 &  1.19$\times10^{-14}$ & 0.41 \\ 
8x & 4XMM J173948.5+095015 &   AGN &    0.43 &      0.22 &  0.15 &        0 & 4.89 &  51.35 &   0.61 &  1.58$\times10^{-14}$ & 0.39 \\ 
\hline\hline
1$^{\S}$ & 2CXO J174025.2+094700 & NS & 0.45 & 0.24 & 0.28 & 0 & 4.71 & 5.7 & 5.28 & 2.89$\times10^{-14}$ & 0.51 \\ 
2 & 2CXO J174011.3+095122 & AGN & 0.44 & 0.12 & 0.92 & 3 & 2.95 & 3.64 & 5.84 & 4.95$\times10^{-14}$ & 0.37 \\ 
3$^{\ddagger}$ & 2CXO J174008.4+095259 & AGN & 0.99 & 0.02 & 26.24 & 2 & 4.40  & 7.67 & 0.93 & 1.72$\times10^{-14}$ & -1.0 \\ 
4$^{\dagger}$ & 2CXO J174001.7+094850 & - & - & - & - & - & 1.37 & 2.41 & 6.84 & 0.55$\times10^{-14}$ & -1.0 \\ 
5 & 2CXO J173959.1+094817 & AGN & 0.934 & 0.06 & 8.91  & 3 & 2.01 & 4.28 & 4.27 & 1.03$\times10^{-14}$ & 0.09 \\ 
6 & 2CXO J174002.1+094721 & AGN & 0.88 & 0.08 & 7.35 & 2 & 1.75 & 6.45 & 4.63 & 4.26$\times10^{-14}$ & 0.46 \\ 
7c & 2CXO J173952.1+094733 & AGN & 0.72 & 0.04 & 6.96 & 4 & 3.86 & 4.61 & 6.02 & 2.48$\times10^{-14}$ & -0.34 \\ 
8c & 2CXO J173957.1+094521 & AGN & 0.85 & 0.10 & 5.35 & 2 & 4.08  & 5.09 & 6.92 & 3.42$\times10^{-14}$ & 0.38 \\ 
9c$^{*}$ & 2CXO J174002.8+094315 & - & - & - & - & 7 & 5.44 & 3.56 & 11.53 & 1.10$\times10^{-14}$ & -1.0 \\ 
10c$^{*}$ &2CXO J174005.4+094323 & - & - & - & - & 8 & 5.24 & 1.82 & 15.27 & 0.71$\times10^{-14}$ & -1.0 \\ 

 \hline
\end{tabular}
\end{center}
} \vspace{-0.2cm}
 \footnotesize
 { Note: The number on the left corresponds to the X-ray sources in Figure \ref{fig:xray_images}. ``x'' and ``c'' symbols mark  sources only appearing in 4XMM-DR13 and CSCv2.1, respectively.
 $^a$Machine learning classification probability. $^b$Classification probability dispersion.
 $^c$ Confidence threshold (see appendix). 
 $^d$Number of potential MW 
 counterparts matched to the X-ray sources. $^e$Separation between the X-ray sources and the center of the LHAASO source. $^f$Detection significance ({\tt  sc\_det\_ml} from the 4XMM-DR13 or {\tt significance} from the CSCv2.1). $^g$Positional uncertainty (PU) ({\tt sc\_poserr} from 4XMM-DR13 or $({\tt error\_ellipse\_r0}^2+{\tt error\_ellipse\_r1}^2)^{1/2}$ from the CSCv2.1). $^h$Observed flux({\tt sc\_ep\_8\_flux} from 4XMM-DR13, 0.2 to 12 keV; or {\tt flux\_aper90\_avg\_b} from CSCv2.1, 0.5-7.0 keV).  
 $^i$Hardness ratio defined in Figure \ref{X_o_flux_HR_XMM}.
 $^{\ddagger}$The source is variable ({\tt sc\_var\_flag} from 4XMM-DR13 or {\tt var\_flag} from CSCv2.1 is True, or the inter-observation probability of being not variable {\tt Var\_Prob} $<0.27\%$ in 4XMM-DR13s \citep{2020A&A...641A.137T}).
 $^{\dagger}$The source is not classified due to the low detection significance. $^{*}$The source is not classified due to the large number of  MW counterparts.
 $^{\S}$The CXO source did not match the MW counterpart despite a larger PU due to its large offset from the nearest MW counterpart.
 }
 \vspace{-0.0cm}
 \label{tab:combine_table}
\end{table*}

\begin{figure*}[h]
\centering
\includegraphics[height=20cm,angle=0]{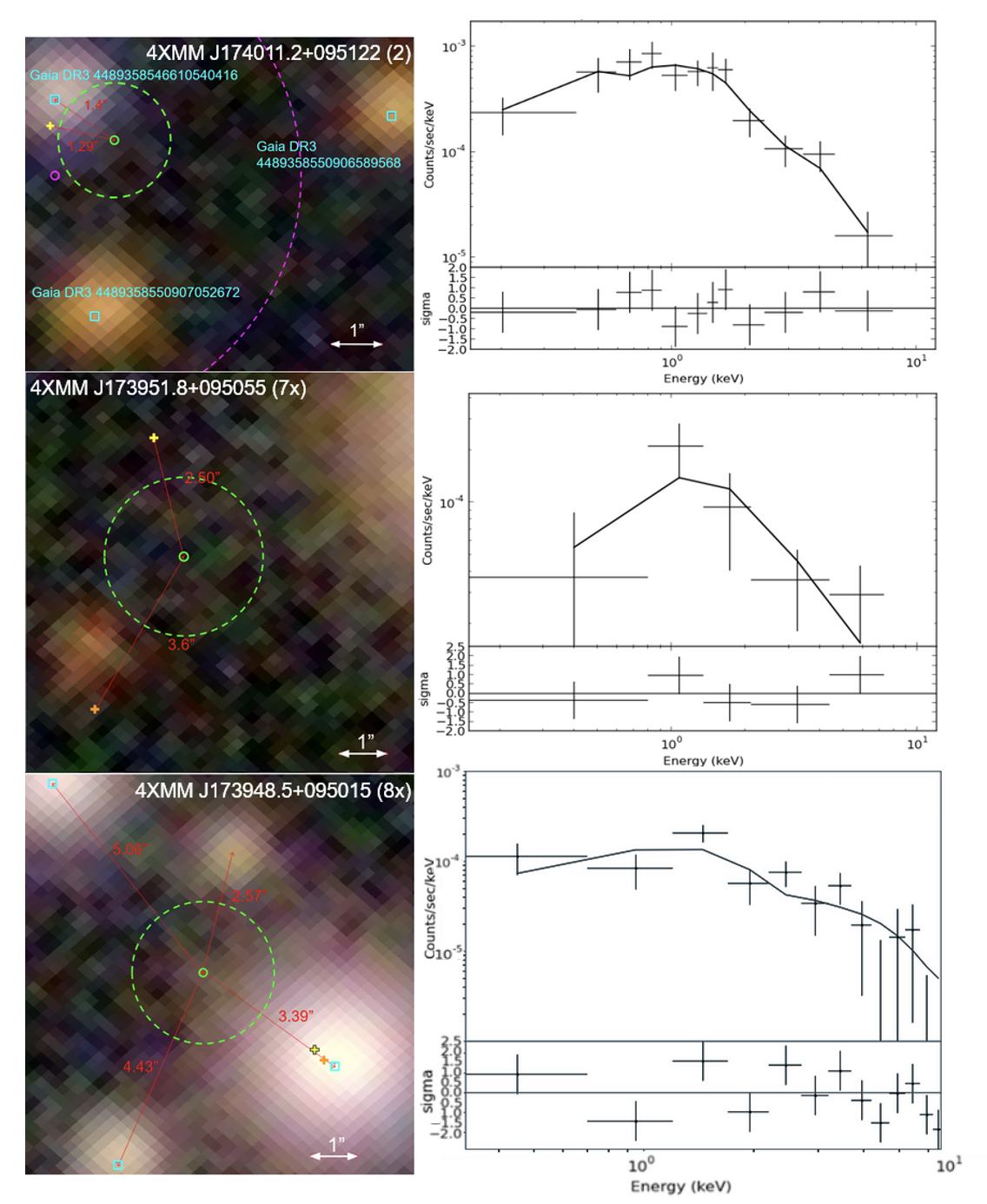}
\caption{\footnotesize
 Pan-STARRS optical images images (left) of the regions around src2 (top), src7x (middle), and src8x (bottom). The {\sl XMM-Newton} sources are labeled with green circles, the CXO sources are labeled with purple circles, the 2-sigma error circles are shown with dash lines respectively ({\tt error\_ellipse\_r0} is used for the CXO error circle). The Gaia sources are labeled with cyan squares, the CatWISE sources are labeled with yellow crosses, and the ALLWISE sources are labeled with orange crosses. The corresponding {\sl XMM-Newton} MOS2 spectra (right).}
\label{fig2}
\end{figure*}

\noindent {\em src2:}  While 4XMM J174011.2+095122 and its CXO counterpart are both classified as a  low-confidence AGN,  
the reasons are likely different. 
The {\sl XMM-Newton} source did not match any MW counterpart due to 
its 95\% confidence error circle being only 1.12$''$ while the nearest counterpart is at 1.29$''$. On the other hand, the CXO source with a 95\% confidence error ellipse of (4.82$''$, 3.29$''$) is matched to two 
distinct counterparts, Gaia DR3 4489358546610540416  and 4489358550907052672. 
Since both the CXO and XMM-Newton positions  are much closer to Gaia DR3 4489358546610540416 (see the top panel in Figure \ref{fig2}),  this Gaia source, and the associated WISE counterpart, are likely to be the true counterparts of the X-ray source and its PU in 4XMM-DR13 may be underestimated.
Moreover, since this Gaia and CatWISE sources 
has no significant proper motion, 
it is likely extragalactic.
The {\sl XMM-Newton} spectrum, available from 4XMM-DR13, fits with an absorbed power-law of $\Gamma \simeq 1.8$ common among  AGN.
Furthermore,
 while the combined classification (for both MW associations) 
 of the CXO source is a CT=0.92 AGN, the 
 association with Gaia DR3 4489358546610540416 
is classified as an AGN with a CT=1.65. Thus, 
src2 is likely 
to be an AGN.

\noindent {\em  src7x:}
4XMM J173951.8+095055 has 95\% confidence 
PU of 1.62$''$ which is much smaller than the distance to the MW closest counterpart. The closest MW source is a  
CatWISE source located  2.5$''$ away and the closest ALLWISE source 
is 3.6$''$ away. In the Pan-STARRS image, we find an optical source that coincides with the ALLWISE source. 
Since these sources are located too far to be associated with the 4XMM source (unless its PU is strongly underestimated), 
 we assume that 
 it has no MW counterparts. 
 A lower limit on the X-ray to optical flux ratio is calculated to be $\approx 1$, using PanSTARRS's mean 5-$\sigma$ sensitivity of $g=23.3$ mag.
According to Figure \ref{X_o_flux_HR_XMM}, this limit is still compatible with an X-ray-faint AGN.
The 
4XMM-DR13 spectrum of src7x (middle panel of Figure \ref{fig2}) 
fits an absorbed power-law 
with 
$\Gamma \simeq 1.1$ which is somewhat hard for AGN  or a pulsar, although not unheard of.
For a Galactic source at $d=1$ kpc (which corresponds to the thickness of the galactic disk), we estimate 
the maximum luminosity of $L_{0.2-12.0 \mathrm{ keV}}=1.4\times10^{30}\ \mathrm{erg} \ \mathrm{s}^{-1} $, which would be compatible with those of old ($>1$ Myr) pulsars (see \citealt{2012ApJ...761..117P}).
However, at such an advanced age a  pulsar is unlikely 
to be producing detectable UHE emission at $d=1$ kpc. Such a hard spectrum is  more typical for young rotation-powered pulsar which should then be  more luminous. 

\noindent {\em src8x:} 
4XMM J173948.5+095015  has 95\% confidence 
PU of 1.49$''$ and 
does not match any MW counterpart (see the  bottom panel of Figure \ref{fig2}), similar to src7x.
The X-ray spectrum 
fits an absorbed power-law model 
with $\Gamma=1.29\pm0.24$. 
The source's X-ray spectrum is harder than those of most middle-aged or old  pulsars. 
However, similar to src7x, the X-ray luminosity would be too low for a young pulsar at $d=1$ kpc or less. 

Finally,  we did not attempt to classify two CXO sources (9c and 10c), lacking {\sl XMM-Newton} coverage, because they matched $>5$ counterparts\footnote{Their PUs are very large due to due to the extremely off-axis placement on the ACIS detector.}. Better  X-ray  localizations are needed to perform meaningful MW classifications with our automated pipeline. 

\section{\label{sec:Fermi_analysis}Fermi LAT data analysis}

The Fermi Gamma-ray Observatory's  Large Area Telescope (LAT) instrument is sensitive to $\gamma$-rays between energies from 20\,MeV to $> 300$\,GeV \citep{atwood_large_2009,atwood_pass_2013,abdollahi_fermi_2020}. In our analysis we include all events within a $r=10^{\circ}$, circular region of interest (ROI) from  LHASSO J1740.5+0948u 
($l = 33.800\degree, b = +20.260\degree$) 
with dates ranging from MJD 54,682 to MJD 60,492 (from 2008 August 4 to 2023 May 28). We chose the energy range between 5\,GeV and 2\,TeV to avoid the larger PSF at lower energies and to minimize the contamination from the nearbypoint-source, 4FGL J1740.5+1005, thought to be associated with PSR J1740+1000, whose pulsed emission has been marginally detected with Fermi LAT \citep{2022MNRAS.513.3113R}. The events were filtered with standard data quality selections ``(DATA$\_$QUAL$>$0)$\&\&$(LAT$\_$CONFIG==1)'', and excluding zenith angles above 90$^{\mathrm o}$. We chose to include both front and back converting events using ``evtype=3'' and ``evclass=128''.

\begin{figure}
\centering
\includegraphics[width=1.0\linewidth]{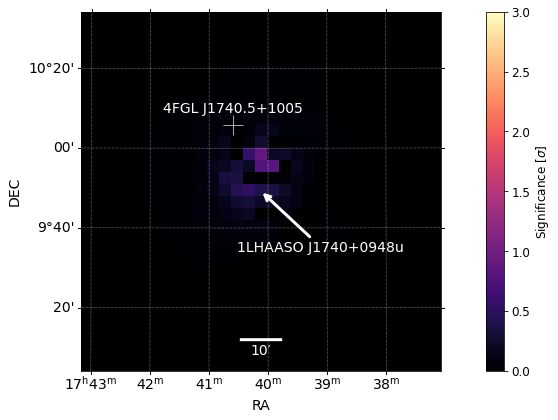}
\caption{Test Statistic (TS) map centered on the source 1LHAASO J1740+0948u from 5--50\,GeV showing the significance of the LHAASO source. Contributions of all other GeV sources within the  region shown are subtracted out. The color scale spans TS values from 0 up to 9 (an approximately $3\sigma$ detection), highlighting regions of potential $\gamma$-ray excess. The white cross marks the catalog position of PSR J1740+1000 and an arrow pointing to the LHAASO position of the \,TeV source.  
}
\label{fig:enter-label}
\end{figure}

The Fermi--LAT events were processed using \texttt{FermiTools}, \citep[version 2.2,][]{atwood_large_2009} and the FermiPy analysis package \citep[version 1.2.0,][]{2017ICRC...35..824W} for Python. All Fermi--LAT analyses use Pass 8 data \citep{atwood_pass_2013,bruel2018fermilatimprovedpass8event} and the 
4FGL-DR4 catalog \citep{Abdollahi_2022} source models 
for all sources within $15^{\circ}$ of LHAASO J1740.5+0948u to ensure all potential sources of the events are accounted for.
We add a point-source at the location of the LHAASO source \citep{2025100802}, modeling its spectrum with a power-law. We use choose to include 4FGL J1740.5+1005 in our background model (to account for the GeV emission from the pulsar) and use the 4FGL-DR4 catalog \citep{2023arXiv230712546B} values for 4FGL J1740.5+1005 position and spectrum.
We then use the maximum likelihood analysis technique  to find the best-fit model of the entire ROI. A test statistic (TS) is associated with every source in the ROI and measures the significance of the presence of the source compared to the absence of the source. Upon fitting,
we find a TS value of only  
2.80
for 1LHAASO J1740+0948u for the entire energy range (5\,GeV -- 2\,TeV), and therefore we do not consider this a significant detection of the LHAASO source.

From this, we measure upper-limits from 4 logarithmically spaced energy bins from 10\,GeV to 2\,TeV. We do not find a statistically significant (TS$>$9)
detection for any energy bins and thus only provide upper-limits for our multi-wavelength modeling. Lastly, we attempted to update the position and localize the point source added to characterize the LHAASO emission, but did not find a significant improvement of the position reported by \citet{2025100802}.

\section{\label{sec:discussion}Discussion}

Below we 
discuss  different scenarios for the origin of TeV emission from LHAASO J1740+0948u. We also discuss the implications for particle acceleration within the scenario where the UHE source is powered by the outflow of energetic particles from PSR J1740+1000, which we  deem the most likely scenario after considering other options. 

\subsection{\label{sec:tail_emission}Emission from pulsar tail}

\begin{figure*}
    \centering
\includegraphics[width=1.0\linewidth]{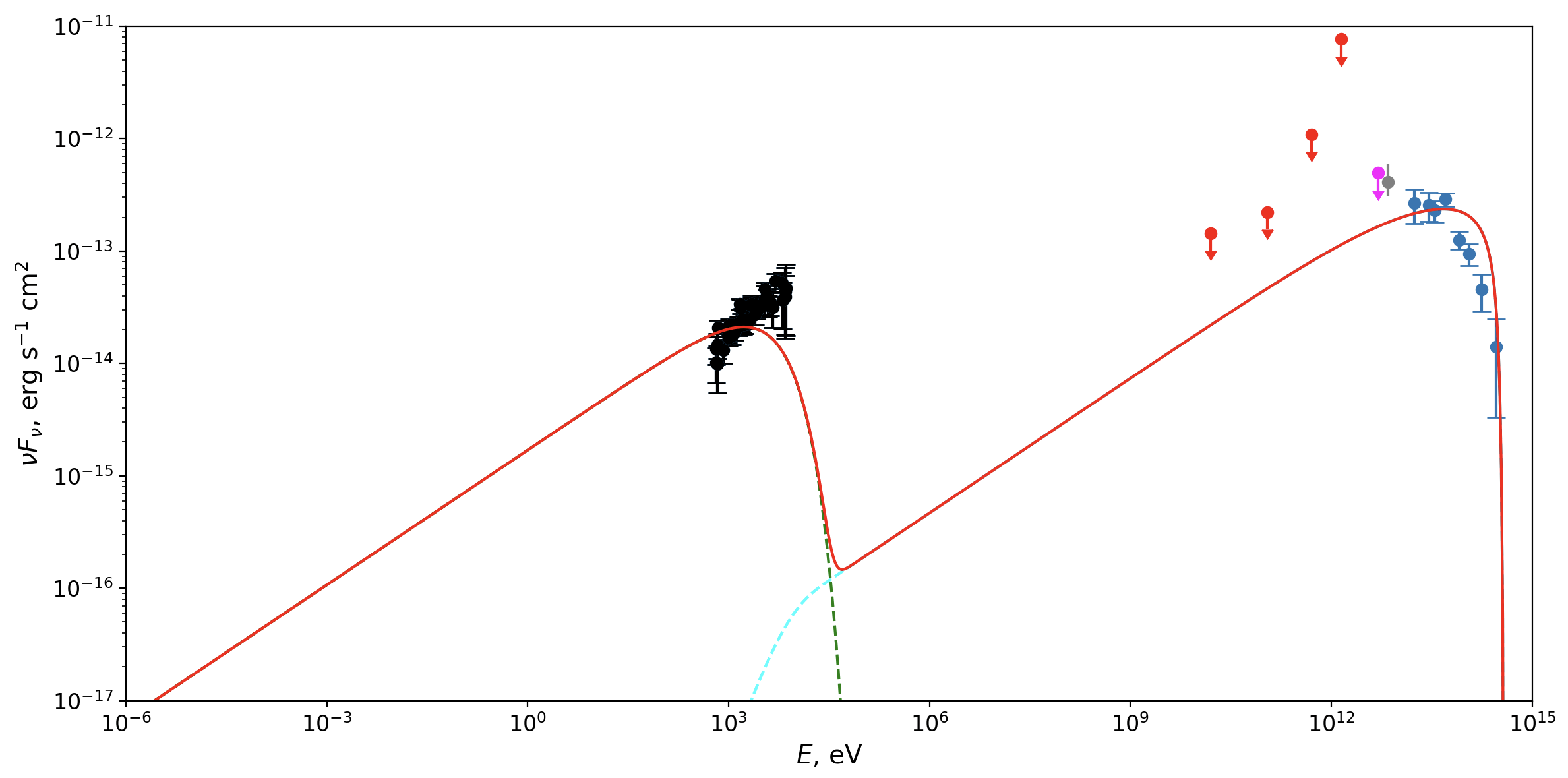}
    \caption{ MW SED for the outflow from J7140 together with SED of LHAASO J1740+0948u. Shown are the data from
XMM-Newton (black; this paper), Fermi-LAT (red upper limits; this paper), VERITAS (magenta upper limt; \citealt{2021ApJ...916..117B}), LHAASO (blue; \citealt{2024ApJS..271...25C}), and HAWC (grey; \citealt{2020ApJ...905...76A}). The red curve shows the spectrum based on the tail model from \cite{2021ApJ...916..117B}  with synchrotron emission
(dashed green), the IC emission  (dashed cyan). The model parameters are: $z_0=6.2\times 10^{17}$ cm, $E_{\rm max}=400$  TeV,  $\beta=-0.3$, $\gamma=1$, $\alpha=-0.7$, $B_0=1~\mu$ G, $v_0=50000$ km s$^{-1}$ (see \citealt{2021ApJ...916..117B} for 
the model details). Note that the X-ray spectrum was extracted from the smaller volume then the extent of the tail from pulsar to the UHE emission cite used for the spatial integration in the model. If some of the very faint X-ray is unaccounted for due to the smaller extraction region, X-ray flux may be slightly higher and spectrum may be somewhat different. }
    \label{fig:mw_sed}
\end{figure*}

Klein-Nishina (KN) corrections are important when considering production of $>100$ TeV $\gamma$-rays by up-scattering CMB photons off the ultra-relativistic pulsar wind electrons.  In 
this regime 
the electron energy can be estimated as $E_e\simeq365 (E_{\gamma}/100~{\rm TeV})^{0.77}$ TeV where $E_{\gamma}$ is the up-scattered photon energy \citep{2021Sci...373..425L}.
Such electrons lose energy via both synchrotron radiation and IC scattering with  the total radiated power   $P_{\rm loss} = P_{\rm syn} + P_{\rm IC, KN}$. 
For the highest energy electrons ($E_e\approx850$ TeV corresponding to $E_{\gamma}=300$ TeV), 
the synchrotron power is 
$P_{\rm syn} = 4 \sigma_T c \gamma^2 U_B /3 
\approx 3.09 \times 10^{-9} (E_e/850~{\rm TeV})^2(B/1~\mu{\rm G})^2~\mathrm{erg\,s}^{-1} $ 
while the IC power in KN regime is $P_{\rm IC, KN} = 4 \sigma_T c \gamma^2 U_{\rm CMB} F_{KN}/3=1.14 \times 10^{-9}~\mathrm{erg\,s}^{-1}$, where $F_{\rm KN} \approx 0.036$ for $E_e=850$ TeV \citep{2005MNRAS.363..954M,2010ApJ...710..236S}. 
The corresponding cooling time  is 
$t_c\approx 10$ kyrs for $B=1~\mu$G but it can be somewhat smaller in the scenario where pulsar wind particles flow within the tail where the field 
is higher closer to the pulsar. However, more detailed modeling of the tail's SED suggests that having a stronger field in any substantial part of the tail will over predict the observed X-ray fluxes (see Fig. \ref{fig:mw_sed}). 
It is unlikely that the magnetic field within the tail is lower than $\approx$1 $\mu$G (see the SED modeling below), therefore the cooling time cannot be much longer than the estimated $\approx$10 kyrs, which
is much shorter than the pulsar's characteristic age of  $\sim 114$ kyr. 
On the other hand, for the plausible transverse pulsar velocity of $v_p=200$ km s$^{-1}$ \citep{2025arXiv250215447C},  it would have traversed the $13'$ distance from the center of the LHAASO source to its current location in $\approx 22$ kyrs. Since the flow speed in the pulsar tail can easily exceed $v_p$ for at least some fraction of the length of the tail, these timescales are compatible. 
In fact, the tail's hard photon index indicates a population of high-energy electrons emitting synchrotron radiation 
in a relatively low magnetic field environment
as they are transported quickly along the tail.

One important implication of the relatively 
small cooling timescale for $E_e=850$ TeV particles is that the pulsar was not much younger  at the time when it produced particles responsible for the observed UHE emission from LHAASO J1740+0948u. This implies that the pulsar's $\dot{E}$ 
at that time was not much larger than it currently is. For the current $\dot{E}$ the potential drop across the polar cap corresponds to electrons gaining an energy of  $E_{\rm pc}=e(\dot{E}/c)^{1/2}=830$  TeV which is very close to the maximum $E_{e}=850$ TeV inferred above from IC mechanism.  This is surprising because $E_{\rm pc}$ is typically considered to be an absolute maximum energy available from pulsar for particle acceleration (e.g., \citealt{2022ApJ...930L...2D,2024arXiv240210912A}), although, \cite{2020A&A...635A.138G} notes that the full vacuum potential drop from
pole to equator is a factor of $R_{\rm LC}/R_{\rm NS}$  larger\
and some of the energy may be available if acceleration occurs in reconnecting regions located in  equatorial plane outside the light cylinder distance, $R_{\rm LC}$. 

While the exact size of the LHAASO source is not known, an upper limit on its angular size\footnote{Note that the limit corresponds to the radius of the region containing 39\% of the source's flux for a point source model fit, at 95\% confidence \citep{2024ApJS..271...25C}.}, $\delta<0.15^{\circ}\approx9'$, is difficult to reconcile with the angular size expected from the often  assumed isotropic Bohm  diffusion,  
$\delta_B=L_B/d=(6Dt_c)^{1/2}d^{-1}$ $=74' (B/1~\mu{\rm G})^{-1/2}(E_e/100~{\rm TeV})^{1/2} (t/10~{\rm kyrs})^{1/2} $ $\times(d/1.2~{\rm kpc})^{-1}$. Note that we used $E_e=100~{\rm TeV} $ in the above estimate because the size of the LHAASO source is not determined by the most extreme photons observed but rather by most common photons and hence can be significantly larger for the highest energy photons.  In fact, the gyration  radius of a 500 TeV particle, $r_g=0.54 (E_e/500~{\rm TeV})(B/1~\mu\rm{G})^{-1}$ pc, corresponds to the angular size of $\delta_g=1.5'$ at d=1.2 kpc. Therefore, if magnetic field is indeed as low as $\simeq1~\mu\rm{G}$, the diffusion approximation is not good and the $\approx500$ TeV particles expected to escape rapidly over a few gyration timescales, $t_g=8(E_e/500~{\rm TeV})(B/1~\mu\rm{G})^{-1}$ yrs. However, the diffusion approximation still may be applicable to particles with $E_e\lesssim100$ TeV. If we assume a more realistic 
diffusion coefficient, $D(E_e)=D_0(E_e/E_{e,0})^{\delta}$, with $D_0=5\times 10^{28}$ cm$^2$ s$^{-1}$ at $E_{e,0}=4$ GeV and $\delta=0.4-0.5$ inferred for local ISM  \citep{2017PhRvD..95h3007Y}, 
we obtain the diffusion size (radius) of tens of degrees.  
This is a known issue for TeV halos, with models of Geminga and B0656+14 halos requiring diffusion suppression by 2-3 orders of magnitude \citep{2024ApJ...974..246A}. 
  \cite{2019MNRAS.488.4074F} argued that fast pulsar motion excludes the suppression of diffusion by scattering from resonant Alfven waves excited in the pulsar wind but argued that slow diffusion may arise from the strong
turbulence  generated by the shock wave of the parent SNR, {\em if the pulsar is still with it}. 
Another  possibility, is a strongly anisotropic diffusion (see, e.g., \citealt{2025ApJ...987...19Y} but also a critique in \citealt{2022PhRvD.106l3033D}), however, even in this case the cross-field diffusion still must be smaller than commonly assumed.   In order to explain the observed picture, an elongated X-ray feature  extending from pulsar toward a fairly compact UHE source, one has to either assume a highly anisotropic diffusion with particles streaming very fast along the magnetic filed lines within the X-ray feature until they reach the UHE site while very slowly diffusing cross-field (a ``pulsar filament scenario'') or an advective transport along most of the X-ray feature's extent being dominant  over the diffusion  which then still needs to be very slow (the tail scenario).  
In either case, particle transport speed can, at least  initially, 
greatly exceed that of the pulsar. In the tail scenario,   the flow is likely to gradually 
slow down with increasing distance from the puslar on angular scales of arcminutes (see \citealt{2008ApJ...684..542K} for more detailed discussion). 
An unrealistically large magnetic field,  $B\simeq 100$~$\mu$G, would be required to confine the  UHE-emitting particles within the $r\lesssim9'$ region, if Bohm diffusion coefficient is used.
With such a high field the UHE emitting region would have been rather bright in X-rays given the observed very high  particle energies.  
An alternative way to decrease the diffusive  source size  is to 
decrease the age of particles within the UHE site to significantly (a factor of 100 reduction would be required for $B=1$~$\mu$G) below the $\tau_c$, which  does not seem to be very plausible either. Although the particles could, in principle, be delivered to the UHE site much quicker than $\tau_c$ via fast  bulk flow in the tail, it would then be unclear why they  abruptly stop there and not be transported any farther from the pulsar unless 
the super-slow diffusion quickly becomes the dominant transport mechanism at $\simeq 13'$ distance from the pulsar. In principle it might  be possible if the ordered flow rapidly becomes very turbulent at that  distance. 

Since the tail scenario is one of the possibilities and there is an already existing model  
\citep{2021ApJ...916..117B}, we use this model  
to calculate the particle SED evolution with the distance from the pulsar and  MW emission from the pulsar tail. 
The model parameters (corresponding to the SED model shown in Figure
\ref{fig:mw_sed}) are (1) the injection particle SED $N(E_e, z_0) = N_0 E_e^{-p}$ with $p=2.4$ (constrained by the slope of the X-ray spectrum; $p=2\Gamma-1$ for synchrotron) with the  maximum particle energy $E_{\rm max}=400$ TeV;  (2)  the exponents, $\beta=-0.3$, $\alpha=-0.7$, and 
$\gamma=1$ (conical tail), for the power-law dependencies  of pulsar
wind  magnetic field $B(z)=B_0(z/z_0)^{\beta}$,  the wind bulk flow speed
$v(z)=v_0(z/z_0)^{\alpha}$, and the tail radius
$r(z)=r_0(z/z_0)^{\gamma}$ on distance from the pulsar, and (3) the  initial 
(at $z=z_0$) values $B_0=1~\mu$ G and $v_0=50,000$ km s$^{-1}$. Of these, $B_0$, $v_0$, $\beta$, $E_{\rm max}$, p, and $N_0$  were varied to match the data as well as possible.
In order to explain
that $E_{\gamma}>100$ TeV emission is the brightest about $\sim13'$ ($\sim 5$ pc) from the pulsar, a rapid flow, which only modestly slows down by that distance, is needed to minimize the IC losses. A low magnetic field for most of the tail is also required   to avoid synchrotron losses. Having the decelerating bulk flow 
within a widening tail (we assume a conical shape guided by X-ray images) ensures a larger number of particles farther away from the pulsar after integrating over the tail diameter. This, in turn, causes IC brightness to peak farther down the tail. On the other hand, having magnetic field increase toward the pulsar explains why X-ray emission is brighter closer to the pulsar. Although these factors are included in the model, we are only able to crudely approximate the measured MW spectrum with the 
model SED 
(see
Figure \ref{fig:mw_sed})
and we have
to set magnetic field to a low value at the base (start) of the tail (taken to be at $z_0=30''$ or 0.2 pc) 
where particles are injected into the tail with initial SED $N(E_e, z_0)\propto E_e^{-p}$. 
We note that although the initial field is already low, the external to the tail ISM field could be even lower because the pulsar is moving outside of the Galactic plane. The fast flow and low magnetic field imply that radiative cooling impact onto particle energies (all the way up to $E_e=800$ TeV) is negligible. Increasing the magnetic field or decreasing the flow speed introduces a large discrepancy between the model and the 
measurements due to the cooling 
causing a very different (from the observed) spectral slope in X-rays and a larger X-ray-to-TeV flux ratio than is observed.
Sensitive radio observations could reveal an even longer extent of the tail 
due to the much longer cooling timescale for particles emitting in the radio. Also, one could expect more sensitive TeV observations with CTA (Cherenkov Telescope Array) to reveal an extension of the TeV source toward the pulsar. 
 We note that the requirements of the very fast flow speed and low magnetic field  occur  naturally within a different, pulsar filament, scenario which we do not attempt to model here but discuss more below.  

Figure \ref{fig:mw_sed} 
shows the MW SED of the pulsar tail and UHE region which we consider to be part of the tail. For this figure we used the X-ray spectrum from {\sl XMM-Newton} described in Section \ref{sec:xray_fits},
the Fermi LAT upper-limits from Section \ref{sec:Fermi_analysis}, the VERITAS upper limits from \cite{2021ApJ...916..117B}, the HAWC detection from \cite{2020ApJ...905...76A}, and the LHAASO detection from \cite{2024ApJS..271...25C}. 
\cite{2021ApJ...916..117B}
computed the model SED for the J1740 tail but, since the paper was written prior to the LHAASO detection of TeV emission, the distance from the pulsar to the TeV emission site and the size of the TeV source were not known.  Moreover, here we performed a more accurate measurements  of the X-ray spectrum of the tail from 
multiple {\sl XMM-Newton}  observations.
\citealt{2025100802} also modeled the SED by fitting the TeV data with an inverse-Compton component, however, they excluded the X-ray band from the fit due to the spatial offset between the LHAASO source and the pulsar tail.
Therefore, we redo the modeling using new observational constraints. 
Guided by the deep {\sl XMM-Newton} EPIC MOS images, 
we 
now assume a more realistic conical tail shape  (unlike the cylindrical one assumed in \citealt{2021ApJ...916..117B}).
We note that the \cite{2021ApJ...916..117B} model has several
approximations and limitations.  The functional forms for $B$, $v$, and
$r$  are assumed to be power-laws rather than obtained by solving,
e.g.,  MHD equations. A more realistic modeling of the flow by, e.g.,
\cite{2018MNRAS.478.2074B}, suggests more complex functional forms which
would require a numerical solution.  Although the tail cross-section
increases, the increase is assumed to be gradual enough to neglect the
transverse (to the tail axis) velocity component. The magnetic field
inside the tail is assumed to be purely toroidal, which is only strictly
possible for a cylindrical tail. This assumption, together with the
magnetic flux freezing valid  for ideal MHD, links one of the
exponents to the others, i.e., $\gamma=-\alpha-\beta$ (see, e.g.,
\citealt{2009ApJ...703..662R}). The \cite{2021ApJ...916..117B} model
also does not take into account any matter
entrainment external to the tail and diffusion transport, which should become
increasingly important at larger distances.

We compute the 
radiation spectrum using the python package {\tt Naima} \citep{2015ICRC...34..922Z}. The emission model is composed of synchrotron and IC\footnote{We only include CMB in our calculations because the source is above the plane and typical values of IR  emission from dust and starlight used for the plane may be inaccurate at that location \citep{2023ApJ...954..200A}. We check that including them does not change the results noticeably.}   components
computed from the electron SED\footnote{We note that equations (3) and (4) in \cite{2021ApJ...916..117B} have  typos in the sign of $\lambda_2$ exponent in their denominators.} given by equation (4) in \cite{2021ApJ...916..117B}. In this model the maximum energy of injected particles 
is strongly constrained  by the drop off of the TeV spectrum to be
$\approx 500$
TeV, 
corresponding to 30\% of the total potential drop across the polar cap.
The  modeling predicts a factor of 2 lower  maximum particle energy than what follows from 
the estimate at the beginning of this section. We attribute this to the approximations \citep{2014ApJ...783..100K} used in the {\tt Naima}'s radiative code \citep{2015ICRC...34..922Z} and the above analytic estimate.  For the model parameters specified in the caption of Figure \ref{fig:mw_sed}, the model spectrum only roughly matches the data,
although it underpredicts the data in the 2-5 keV range. We were not able to obtain a significantly better fit with this simplistic tail model, which, nonetheless, is more appropriate in this case than the often used isotropic one-zone model. 
 
\begin{figure*}
    \centering
\includegraphics[width=0.8\linewidth]{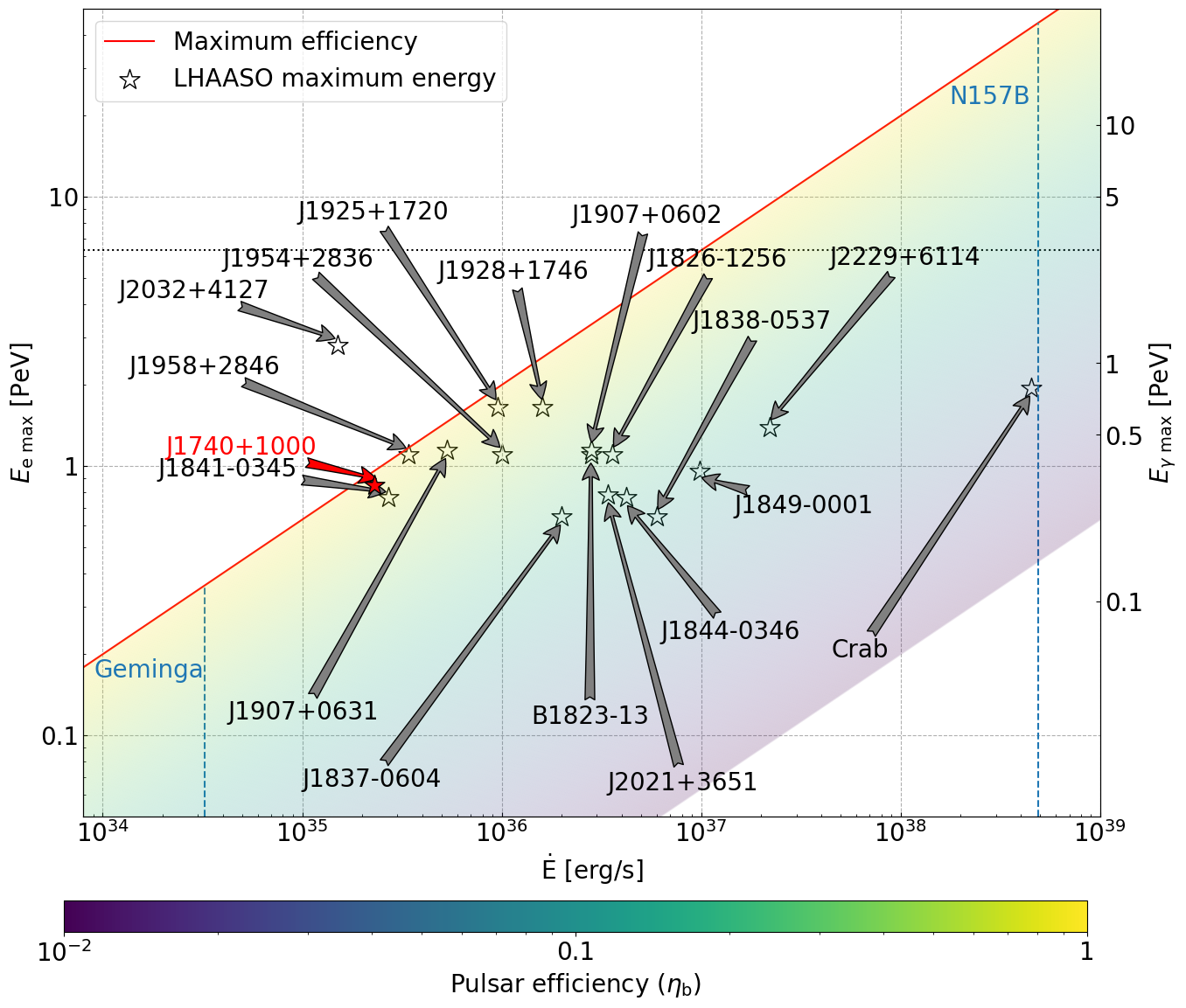}
    \caption{Maximum electron energy inferred from the LHAASO spectra of UHE sources vs.\ spin-down power of potentially associated pulsars. PSR J1740+1000 is shown by the red filled star. Adapted form \cite{2022ApJ...930L...2D}.}
    \label{fig:popul}
\end{figure*}

As one can see from Figure \ref{fig:popul}, assuming the TeV emission is indeed produced by the PWN particles, J1740 
could be the most extreme accelerator (closest to the red line) among isolated pulsars\footnote{PSR J2032+4127 is part of high-mass $\gamma$-ray binary TeV J2032+4130/MT91 213 where a different accleration process related to the intra-binary shock may be dominant.}. Although PSR J1958+2836 is only a slightly less extreme, it is located  inside the Galactic plane in a much more complicated region which also includes an SNR. Therefore, after ruling out other possibilities (see Section \ref{sec:tev_sources}),  PSR J1740+1000 is
the cleanest case demonstrating that pulsar wind particles are able be accelerated to a large fraction of energy associated with the full polar cap potential drop. 

We note that 
the hydrodynamical tail model used above may be 
inappropriate and, instead, the extended X-ray feature could turn out to be  another case of pulsar filament similar to those of  Lighthouse  \citep{2023ApJ...950..177K} and Guitar  \citep{2022ApJ...939...70D} nebulae. However, for the chosen tail model parameters  (see Figure \ref{fig:mw_sed}) the calculated  SED may turn out to be roughly  similar  to the one  expected for particles undergoing a rapid anisotropic diffusion along the ordered ISM field (i.e.\ pulsar filament scenario).   The nature of these recently discovered new kind of pulsar outflow is not fully understood \citep{2024A&A...684L...1O}. However, in the  popular scenario (proposed by \citealt{2008A&A...490L...3B}) the particles that escape are the most energetic ones and hence  should  produce UHE emission. From the recent analysis of the Lighthouse filament \citep{2023ApJ...950..177K} it follows that particles can be streaming away from the pulsar with relativistic speed along ISM magnetic field lines up to some distance where they become trapped due to changing geometry of the ISM magnetic field. 
In case of the Lighthouse PWN, the spectrum of the filament is much harder than the pulsar tail spectrum. J1740 feature's spectrum is similarly hard and harder than spectra of most pulsar tails with X-ray spectral slope measurements. 
Unlike pulsar tails, for the pulsar filament scenario no bright emission in the radio is expected from the X-ray and TeV emitting regions.   Hence, sensitive radio observations can be used to differentiate between the two scenarios.  Also, measuring the pulsar proper motion (velocity)  would provide an additional way to discriminate because the tail will be behind the pulsar while pulsar filaments are typically seen at large angles with respect to proper motion direction. 

\subsection{\label{sec:tev_sources}Alternative sources of TeV emission}
\label{sec: Alt TeV emission}
As an alternative  to pulsar J1740, we also consider a possibility that some other source of energetic particles could be responsible for the observed UHE emission. We do not find any evidence for GeV emission from the UHE site.   We discuss the  X-ray sources detected within the LHAASO J1740+0948u region   in more detail below.  

In principle, LHAASO source emission could be produced by a blazar since blazars are known  to be TeV sources, and, in some cases, are  detected above 100 TeV (see \citealt{
2000NewA....5..377A}). However,
in such a scenario 
the blazar 
would need to be relatively nearby. This is because the observed $\gtrsim$100 TeV photons will be converted to $e^+/e^-$ pairs due to interactions with low-energy extragalactic background light photons via the Breit-Wheeler (BW) process  (see, e.g., \citealt{2009herb.book.....D}). For $E_{\gamma}=100$ TeV photons the BW horizon, defined by requiring optical depth $\tau_{BW}\approx1$
corresponds to redshift $z=0.04$ (see, e.g., Figure 3 in \citealt{2023EPJC...83..192L}) or luminosity distance\footnote{ see https://www.astro.ucla.edu/$~$wright/CosmoCalc.html} of $\approx 180$ Mpc. 

Since 4XMM J174001.8+094849 (src4), confidently  classified as an AGN by MUWCLASS, is the only X-ray source with radio counterparts in NVSS  \citep{1998AJ....115.1693C}, RACS \citep{2020PASA...37...48M}, and VLASS \citep{2020PASP..132c5001L}, and a nearby blazar expected to be bright in radio, 
we only consider this source as a potential blazar candidate. 
NVSS 174001+094850, the radio counterpart to src4, has an integrated 1.4 GHz flux density of 7.1 mJy, which corresponds to $L_{1.4 \mathrm{GHz}}=1.64\times10^{37}$ $ \mathrm{erg}\ \mathrm{s}^{-1}$ at the maximum plausible  $d=180$ Mpc. It is resolved into two\footnote{VLASS1QLCIR J174001.45+094856.0 and VLASS1QLCIR J174001.90+094850.8} different sources in the VLASS catalog \citep{2021ApJS..255...30G},
of which VLASS1QLCIR J174001.90+094850.8 is more consistent with X-ray positions from multiple {\sl XMM-Newton} observations. It has a flux of 1.5 mJy in the 2-4 GHz range, which corresponds to $L_{2-4 \mathrm{GHz}}=9.25\times10^{36}$ $\mathrm{erg}\ \mathrm{s}^{-1}$. 
 For 4XMM J174001.8+094849, the X-ray luminosity at 
the maximum possible distance is
$L_{1 \mathrm{keV}}=2.08\times10^{40}$ $ \mathrm{erg}\ \mathrm{s}^{-1}$. 
\citealt{2016ApJS..224...26M} performed a comprehensive study of 
blazars from radio through gamma-rays, which shows a minimum X-ray luminosity of $L_{1 \mathrm{keV}} \approx 10^{42}$ $ \mathrm{erg}\ \mathrm{s}^{-1}$ and a minimum radio luminosity of $L_{1.4 \mathrm{GHz}}\approx 10^{39}$ $ \mathrm{erg}\ \mathrm{s}^{-1}$. Compared the X-ray and radio luminosity of src4 against this well-known sample of blazars, 
src4 is well below the expected X-ray and radio luminosity for a blazar at $\approx 180$ Mpc.
We conclude that src4, and any other X-ray source classified as AGN but without a detected radio counterpart, is unlikely to be a sufficiently close-by blazar 
responsible for the LHAASO emission. 

Given the classification results discussed in Section \ref{sec:xray_sources}, it is unlikely that any of the 10 unique X-ray sources classified by MUWCLASS are a pulsar. However, 2CXO J174002.8+094315 and 2CXO J174005.4+094323 were excluded from the classification pipeline due to their large PUs causing them to match  to a large number of potential MW counterparts (or they may  lack a counterpart). 2CXO J174002.8+094315 has a hardness ratio $HR=(F_{1.2-2 \mathrm{ keV}}-F_{0.5-1.2 \mathrm{ keV}})/(F_{1.2-2 \mathrm{ keV}}+F_{0.5-1.2 \mathrm{ keV}})=-0.36$ and a broadband flux $F_{0.5-7 \mathrm{ keV}}=7.1\times10^{-15}\ \mathrm{erg} \ \mathrm{s}^{-1} \ \mathrm{cm}^{-2}$, which corresponds to
$L_{0.5-7 \mathrm{ keV}}=8.5\times10^{29}\ \mathrm{erg} \ \mathrm{s}^{-1} $
at the maximum plausible distance of 1 kpc.
2CXO J174005.4+094323 has a hardness ratio $HR=-0.09$ and a broadband flux $F_{0.5-7.0 \mathrm{ keV}}=1.1\times10^{-14}\ \mathrm{erg} \ \mathrm{s}^{-1} \ \mathrm{cm}^{-2}$, which corresponding to an $L_{0.5-7 \mathrm{ keV}}=1.32\times10^{30}\ \mathrm{erg} \ \mathrm{s}^{-1} $. 
The low X-ray luminosity values  indicate that, even if one or both  CXO sources are pulsars, they would be old pulsars that are unlikely to be capable of producing UHE TeV emission. 
Furthermore, there are only 8 out-of-plane ($>|$10$|^{\circ}$) galactic ($<$10 kpc) pulsars with age $<10^{6}$ yrs. Of these, the closest two pulsars have a separation of $\approx$3$^{\circ}$, according to the ATNF catalog \citep{2005AJ....129.1993M}, while the two CXO sources have a separation of $\approx$18$’$ from PSR J1740+1000. As such it is unlikely that either of these two CXO sources is a pulsar.

To conclude, we believe none of the X-ray sources detected within the PU of 1LHAASO J1740+0948u 
are a promising counterpart.

\section{\label{sec:summary}Summary}

After careful analysis of X-ray sources in the region of UHE source 1LHAASO J1740+0948u, we do not find any plausible candidates other than PSR J1740+1000 located  13$'$ north from the UHE source. 
 Additional evidence for association is the X-ray  feature extending from the pulsar towards the UHE source which could be either the pulsar tail or a beam of most energetic particles escaping into the ISM from the apex of the bow-shock 
 (pulsar filament). For the IC emission  mechanism, which is the only plausible scenario at high Galactic latitude of this source, the    electron energies must reach $\approx 800$ TeV which is nearly equal to the current full potential drop across the polar cap. We argue that since the particles are unlikely to be produced long ago (they would have cooled otherwise, unless $B<1$ $\mu$G), the particle acceleration mechanism operating in the wind must be capable of utilizing  a substantial fraction of the  polar cap potential drop. The only other alternative would be some sort of particle re-acceleration operating far from the pulsar which does not seem plausible 
 given the weak magnetic field.

\begin{acknowledgements}
We thank George Pavlov for insightful discussions  and  Emma de Ona Wilhelmi for help with Figure \ref{fig:popul}. 
Support for this work was provided by the National Aeronautics and Space Administration (NASA)
through
Award Numbers 80NSSC18K0636 and 80NSSC22K1575.
N.K. and J.H. acknowledge support from NASA under award number 80GSFC24M0006.

\end{acknowledgements}

 \bibliographystyle{aasjournal}
 \bibliography{references.bib}

\begin{thebibliography}{}
\expandafter\ifx\csname natexlab\endcsname\relax\def\natexlab#1{#1}\fi
\providecommand{\url}[1]{\href{#1}{#1}}
\providecommand{\dodoi}[1]{doi:~\href{http://doi.org/#1}{\nolinkurl{#1}}}
\providecommand{\doeprint}[1]{\href{http://ascl.net/#1}{\nolinkurl{http://ascl.net/#1}}}
\providecommand{\doarXiv}[1]{\href{https://arxiv.org/abs/#1}{\nolinkurl{https://arxiv.org/abs/#1}}}

\bibitem[{Abdollahi {et~al.}(2020)Abdollahi, Acero, Ackermann, Ajello, Atwood, Axelsson, Baldini, Ballet, Barbiellini, Bastieri, Gonzalez, Bellazzini, Berretta, Bissaldi, Blandford, Bloom, Bonino, Bottacini, Brandt, Bregeon, Bruel, Buehler, Burnett, Buson, Cameron, Caputo, Caraveo, Casandjian, Castro, Cavazzuti, Charles, Chaty, Chen, Cheung, Chiaro, Ciprini, Cohen-Tanugi, Cominsky, Coronado-Blázquez, Costantin, Cuoco, Cutini, D’Ammando, DeKlotz, Luque, Palma, Desai, Digel, Lalla, Mauro, Venere, Domínguez, Dumora, Dirirsa, Fegan, Ferrara, Franckowiak, Fukazawa, Funk, Fusco, Gargano, Gasparrini, Giglietto, Giommi, Giordano, Giroletti, Glanzman, Green, Grenier, Griffin, Grondin, Grove, Guiriec, Harding, Hayashi, Hays, Hewitt, Horan, Jóhannesson, Johnson, Kamae, Kerr, Kocevski, Kovac’evic’, Kuss, Landriu, Larsson, Latronico, Lemoine-Goumard, Li, Liodakis, Longo, Loparco, Lott, Lovellette, Lubrano, Madejski, Maldera, Malyshev, Manfreda, Marchesini, Marcotulli, Martí-Devesa, Martin, Massaro, Mazziotta,
  McEnery, Mereu, Meyer, Michelson, Mirabal, Mizuno, Monzani, Morselli, Moskalenko, Negro, Nuss, Ojha, Omodei, Orienti, Orlando, Ormes, Palatiello, Paliya, Paneque, Pei, Peña-Herazo, Perkins, Persic, Pesce-Rollins, Petrosian, Petrov, Piron, Poon, Porter, Principe, Rainò, Rando, Razzano, Razzaque, Reimer, Reimer, Remy, Reposeur, Romani, Parkinson, Schinzel, Serini, Sgrò, Siskind, Smith, Spandre, Spinelli, Strong, Suson, Tajima, Takahashi, Tak, Thayer, Thompson, Tibaldo, Torres, Torresi, Valverde, Klaveren, Zyl, Wood, Yassine, \& Zaharijas}]{abdollahi_fermi_2020}
Abdollahi, S., Acero, F., Ackermann, M., {et~al.} 2020, \apjs, 247, 33, \dodoi{10.3847/1538-4365/ab6bcb}

\bibitem[{Abdollahi {et~al.}(2022)Abdollahi, Acero, Baldini, Ballet, Bastieri, Bellazzini, Berenji, Berretta, Bissaldi, Blandford, Bloom, Bonino, Brill, Britto, Bruel, Burnett, Buson, Cameron, Caputo, Caraveo, Castro, Chaty, Cheung, Chiaro, Cibrario, Ciprini, Coronado-Blázquez, Crnogorcevic, Cutini, D’Ammando, De~Gaetano, Digel, Di~Lalla, Dirirsa, Di~Venere, Domínguez, Fallah~Ramazani, Fegan, Ferrara, Fiori, Fleischhack, Franckowiak, Fukazawa, Funk, Fusco, Galanti, Gammaldi, Gargano, Garrappa, Gasparrini, Giacchino, Giglietto, Giordano, Giroletti, Glanzman, Green, Grenier, Grondin, Guillemot, Guiriec, Gustafsson, Harding, Hays, Hewitt, Horan, Hou, Jóhannesson, Karwin, Kayanoki, Kerr, Kuss, Landriu, Larsson, Latronico, Lemoine-Goumard, Li, Liodakis, Longo, Loparco, Lott, Lubrano, Maldera, Malyshev, Manfreda, Martí-Devesa, Mazziotta, Mereu, Meyer, Michelson, Mirabal, Mitthumsiri, Mizuno, Moiseev, Monzani, Morselli, Moskalenko, Negro, Nuss, Omodei, Orienti, Orlando, Paneque, Pei, Perkins, Persic,
  Pesce-Rollins, Petrosian, Pillera, Poon, Porter, Principe, Rainò, Rando, Rani, Razzano, Razzaque, Reimer, Reimer, Reposeur, Sánchez-Conde, Saz~Parkinson, Scotton, Serini, Sgrò, Siskind, Smith, Spandre, Spinelli, Sueoka, Suson, Tajima, Tak, Thayer, Thompson, Torres, Troja, Valverde, Wood, \& Zaharijas}]{Abdollahi_2022}
Abdollahi, S., Acero, F., Baldini, L., {et~al.} 2022, The Astrophysical Journal Supplement Series, 260, 53, \dodoi{10.3847/1538-4365/ac6751}

\bibitem[{{Aharonian}(2000)}]{2000NewA....5..377A}
{Aharonian}, F.~A. 2000, \na, 5, 377, \dodoi{10.1016/S1384-1076(00)00039-7}

\bibitem[{{Albert} {et~al.}(2020){Albert}, {Alfaro}, {Alvarez}, {Camacho}, {Arteaga-Vel{\'a}zquez}, {Arunbabu}, {Avila Rojas}, {Ayala Solares}, {Baghmanyan}, {Belmont-Moreno}, {BenZvi}, {Brisbois}, {Caballero-Mora}, {Capistr{\'a}n}, {Carrami{\~n}ana}, {Casanova}, {Cotti}, {Couti{\~n}o de Le{\'o}n}, {De la Fuente}, {Diaz Hernandez}, {Diaz-Cruz}, {Dingus}, {DuVernois}, {Durocher}, {D{\'\i}az-V{\'e}lez}, {Ellsworth}, {Engel}, {Espinoza}, {Fan}, {Fang}, {Alonso}, {Fleischhack}, {Fraija}, {Galv{\'a}n-G{\'a}mez}, {Garcia}, {Garc{\'\i}a-Gonz{\'a}lez}, {Garfias}, {Giacinti}, {Gonz{\'a}lez}, {Goodman}, {Harding}, {Hernandez}, {Hinton}, {Hona}, {Huang}, {Hueyotl-Zahuantitla}, {H{\"u}ntemeyer}, {Iriarte}, {Jardin-Blicq}, {Joshi}, {Kieda}, {Lara}, {Lee}, {Le{\'o}n Vargas}, {Linnemann}, {Longinotti}, {Luis-Raya}, {Lundeen}, {L{\'o}pez-Coto}, {Malone}, {Marandon}, {Martinez}, {Martinez-Castellanos}, {Mart{\'\i}nez-Castro}, {Matthews}, {Miranda-Romagnoli}, {Morales-Soto}, {Moreno}, {Mostaf{\'a}}, {Nayerhoda}, {Nellen},
  {Newbold}, {Nisa}, {Noriega-Papaqui}, {Olivera-Nieto}, {Omodei}, {Peisker}, {P{\'e}rez Araujo}, {P{\'e}rez-P{\'e}rez}, {Ren}, {Rho}, {Rivi{\`e}re}, {Rosa-Gonz{\'a}lez}, {Ruiz-Velasco}, {Salazar}, {Salesa Greus}, {Sandoval}, {Schneider}, {Schoorlemmer}, {Serna}, {Sinnis}, {Smith}, {Springer}, {Surajbali}, {Tollefson}, {Torres}, {Torres-Escobedo}, {Ukwatta}, {Ure{\~n}a-Mena}, {Weisgarber}, {Werner}, {Willox}, {Zepeda}, {Zhou}, {de Le{\'o}n}, {{\'A}lvarez}, \& {HAWC Collaboration}}]{2020ApJ...905...76A}
{Albert}, A., {Alfaro}, R., {Alvarez}, C., {et~al.} 2020, \apj, 905, 76, \dodoi{10.3847/1538-4357/abc2d8}

\bibitem[{{Albert} {et~al.}(2024){Albert}, {Alfaro}, {Alvarez}, {Arteaga-Vel{\'a}zquez}, {Avila Rojas}, {Ayala Solares}, {Babu}, {Belmont-Moreno}, {Bernal}, {Caballero-Mora}, {Capistr{\'a}n}, {Carrami{\~n}ana}, {Casanova}, {Cotti}, {Cotzomi}, {Couti{\~n}o de Le{\'o}n}, {de la Fuente}, {Depaoli}, {Di Lalla}, {Diaz Hernandez}, {Dingus}, {DuVernois}, {Durocher}, {D{\'\i}az-V{\'e}lez}, {Engel}, {Espinoza}, {Fan}, {Fang}, {Fraija}, {Garc{\'\i}a-Gonz{\'a}lez}, {Garfias}, {Goksu}, {Gonz{\'a}lez}, {Goodman}, {Groetsch}, {Harding}, {Hern{\'a}ndez-Cadena}, {Herzog}, {H{\"u}ntemeyer}, {Huang}, {Hueyotl-Zahuantitla}, {Iriarte}, {Joshi}, {Kaufmann}, {Kieda}, {Lara}, {Lee}, {Lee}, {Le{\'o}n Vargas}, {Linnemann}, {Longinotti}, {Luis-Raya}, {Malone}, {Martinez}, {Mart{\'\i}nez-Castro}, {Matthews}, {Miranda-Romagnoli}, {Montes}, {Morales-Soto}, {Moreno}, {Mostaf{\'a}}, {Nayerhoda}, {Nellen}, {Noriega-Papaqui}, {Olivera-Nieto}, {Omodei}, {P{\'e}rez Araujo}, {P{\'e}rez-P{\'e}rez}, {Rho}, {Rosa-Gonz{\'a}lez}, {Salazar},
  {Salazar-Gallegos}, {Sandoval}, {Schneider}, {Schwefer}, {Serna-Franco}, {Son}, {Springer}, {Tibolla}, {Tollefson}, {Torres}, {Torres-Escobedo}, {Turner}, {Urea-Mena}, {Varela}, {Villase{\~n}or}, {Wang}, {Watson}, {Willox}, {Wu}, {Yun-C{\'a}rcamo}, {Zhou}, {de Le{\'o}n}, \& {Di Mauro}}]{2024ApJ...974..246A}
---. 2024, \apj, 974, 246, \dodoi{10.3847/1538-4357/ad738e}

\bibitem[{{Amato}(2024)}]{2024arXiv240210912A}
{Amato}, E. 2024, arXiv e-prints, arXiv:2402.10912, \dodoi{10.48550/arXiv.2402.10912}

\bibitem[{{Amenomori} {et~al.}(2023){Amenomori}, {Asano}, {Bao}, {Bi}, {Chen}, {Chen}, {Chen}, {Chen}, {Chen}, {Cirennima}, {Cui}, {Danzengluobu}, {Ding}, {Fang}, {Fang}, {Feng}, {Feng}, {Feng}, {Gao}, {Gomi}, {Gou}, {Guo}, {Guo}, {Hayashi}, {He}, {He}, {Hibino}, {Hotta}, {Hu}, {Hu}, {Hu}, {Huang}, {Jia}, {Jiang}, {Jiang}, {Jin}, {Kasahara}, {Katayose}, {Kato}, {Kato}, {Kawahara}, {Kawashima}, {Kawata}, {Kozai}, {Kurashige}, {Labaciren}, {Le}, {Li}, {Li}, {Li}, {Li}, {Lin}, {Liu}, {Liu}, {Liu}, {Liu}, {Liu}, {Liu}, {Lu}, {Meng}, {Meng}, {Munakata}, {Nagaya}, {Nakamura}, {Nakazawa}, {Nanjo}, {Ning}, {Nishizawa}, {Noguchi}, {Ohnishi}, {Okukawa}, {Ozawa}, {Qian}, {Qian}, {Qu}, {Saito}, {Sakakibara}, {Sakata}, {Sako}, {Sako}, {Sasaki}, {Shao}, {Shibata}, {Shiomi}, {Sugimoto}, {Takano}, {Takita}, {Tan}, {Tateyama}, {Torii}, {Tsuchiya}, {Udo}, {Wang}, {Wang}, {Wang}, {Wangdui}, {Wu}, {Wu}, {Xu}, {Xue}, {Yang}, {Yao}, {Yin}, {Yokoe}, {Yu}, {Yuan}, {Zhai}, {Zhang}, {Zhang}, {Zhang}, {Zhang}, {Zhang}, {Zhang},
  {Zhang}, {Zhao}, {Zhaxisangzhu}, {Zhou}, \& {Zou}}]{2023ApJ...954..200A}
{Amenomori}, M., {Asano}, S., {Bao}, Y.~W., {et~al.} 2023, \apj, 954, 200, \dodoi{10.3847/1538-4357/acebce}

\bibitem[{{Arnaud}(1996)}]{1996ASPC..101...17A}
{Arnaud}, K.~A. 1996, in Astronomical Society of the Pacific Conference Series, Vol. 101, Astronomical Data Analysis Software and Systems V, ed. G.~H. {Jacoby} \& J.~{Barnes}, 17

\bibitem[{Atwood {et~al.}(2013)Atwood, Albert, Baldini, Tinivella, Bregeon, Pesce-Rollins, Sgrò, Bruel, Charles, Drlica-Wagner, Franckowiak, Jogler, Rochester, Usher, Wood, Cohen-Tanugi, \& Zimmer}]{atwood_pass_2013}
Atwood, W., Albert, A., Baldini, L., {et~al.} 2013.
\newblock \url{https://arxiv.org/abs/1303.3514v1}

\bibitem[{Atwood {et~al.}(2009)Atwood, Abdo, Ackermann, Althouse, Anderson, Axelsson, Baldini, Ballet, Band, Barbiellini, Bartelt, Bastieri, Baughman, Bechtol, Bédérède, Bellardi, Bellazzini, Berenji, Bignami, Bisello, Bissaldi, Blandford, Bloom, Bogart, Bonamente, Bonnell, Borgland, Bouvier, Bregeon, Brez, Brigida, Bruel, Burnett, Busetto, Caliandro, Cameron, Caraveo, Carius, Carlson, Casandjian, Cavazzuti, Ceccanti, Cecchi, Charles, Chekhtman, Cheung, Chiang, Chipaux, Cillis, Ciprini, Claus, Cohen-Tanugi, Condamoor, Conrad, Corbet, Corucci, Costamante, Cutini, Davis, Decotigny, DeKlotz, Dermer, de~Angelis, Digel, do~Couto~e Silva, Drell, Dubois, Dumora, Edmonds, Fabiani, Farnier, Favuzzi, Flath, Fleury, Focke, Funk, Fusco, Gargano, Gasparrini, Gehrels, Gentit, Germani, Giebels, Giglietto, Giommi, Giordano, Glanzman, Godfrey, Grenier, Grondin, Grove, Guillemot, Guiriec, Haller, Harding, Hart, Hays, Healey, Hirayama, Hjalmarsdotter, Horn, Hughes, Jóhannesson, Johansson, Johnson, Johnson, Johnson, Johnson,
  Kamae, Katagiri, Kataoka, Kavelaars, Kawai, Kelly, Kerr, Klamra, Knödlseder, Kocian, Komin, Kuehn, Kuss, Landriu, Latronico, Lee, Lee, Lemoine-Goumard, Lionetto, Longo, Loparco, Lott, Lovellette, Lubrano, Madejski, Makeev, Marangelli, Massai, Mazziotta, McEnery, Menon, Meurer, Michelson, Minuti, Mirizzi, Mitthumsiri, Mizuno, Moiseev, Monte, Monzani, Moretti, Morselli, Moskalenko, Murgia, Nakamori, Nishino, Nolan, Norris, Nuss, Ohno, Ohsugi, Omodei, Orlando, Ormes, Paccagnella, Paneque, Panetta, Parent, Pearce, Pepe, Perazzo, Pesce-Rollins, Picozza, Pieri, Pinchera, Piron, Porter, Poupard, Rainò, Rando, Rapposelli, Razzano, Reimer, Reimer, Reposeur, Reyes, Ritz, Rochester, Rodriguez, Romani, Roth, Russell, Ryde, Sabatini, Sadrozinski, Sanchez, Sander, Sapozhnikov, Parkinson, Scargle, Schalk, Scolieri, Sgrò, Share, Shaw, Shimokawabe, Shrader, Sierpowska-Bartosik, Siskind, Smith, Smith, Spandre, Spinelli, Starck, Stephens, Strickman, Strong, Suson, Tajima, Takahashi, Takahashi, Tanaka, Tenze, Tether,
  Thayer, Thayer, Thompson, Tibaldo, Tibolla, Torres, Tosti, Tramacere, Turri, Usher, Vilchez, Vitale, Wang, Watters, Winer, Wood, Ylinen, \& Ziegler}]{atwood_large_2009}
Atwood, W.~B., Abdo, A.~A., Ackermann, M., {et~al.} 2009, \apj, 697, 1071, \dodoi{10.1088/0004-637X/697/2/1071}

\bibitem[{{Ballet} {et~al.}(2023){Ballet}, {Bruel}, {Burnett}, {Lott}, \& {The Fermi-LAT collaboration}}]{2023arXiv230712546B}
{Ballet}, J., {Bruel}, P., {Burnett}, T.~H., {Lott}, B., \& {The Fermi-LAT collaboration}. 2023, arXiv e-prints, arXiv:2307.12546, \dodoi{10.48550/arXiv.2307.12546}

\bibitem[{{Bandiera}(2008)}]{2008A&A...490L...3B}
{Bandiera}, R. 2008, \aap, 490, L3, \dodoi{10.1051/0004-6361:200810666}

\bibitem[{{Benbow} {et~al.}(2021){Benbow}, {Brill}, {Buckley}, {Capasso}, {Chromey}, {Errando}, {Falcone}, {Farrell}, {Feng}, {Finley}, {Foote}, {Fortson}, {Furniss}, {Gent}, {Giuri}, {Hanna}, {Hassan}, {Hervet}, {Holder}, {Hughes}, {Humensky}, {Jin}, {Kaaret}, {Kargaltsev}, {Kertzman}, {Kieda}, {Klingler}, {Kumar}, {Lang}, {Lundy}, {Maier}, {McGrath}, {Moriarty}, {Mukherjee}, {Nieto}, {Nievas-Rosillo}, {O'Brien}, {Ong}, {Otte}, {Patel}, {Pfrang}, {Pohl}, {Prado}, {Quinn}, {Ragan}, {Reynolds}, {Ribeiro}, {Richards}, {Roache}, {Ryan}, {Santander}, {Sembroski}, {Shang}, {Volkov}, {Wakely}, {Weinstein}, {Wilcox}, \& {Williams}}]{2021ApJ...916..117B}
{Benbow}, W., {Brill}, A., {Buckley}, J.~H., {et~al.} 2021, \apj, 916, 117, \dodoi{10.3847/1538-4357/ac05b9}

\bibitem[{Bordas {et~al.}(2009)Bordas, Bosch-Ramon, Paredes, \& Perucho}]{Bordas2009}
Bordas, P., Bosch-Ramon, V., Paredes, J.~M., \& Perucho, M. 2009, Astronomy \& Astrophysics, 497, 325, \dodoi{10.1051/0004-6361/200811539}

\bibitem[{Bruel {et~al.}(2018)Bruel, Burnett, Digel, Johannesson, Omodei, \& Wood}]{bruel2018fermilatimprovedpass8event}
Bruel, P., Burnett, T.~H., Digel, S.~W., {et~al.} 2018.
\newblock \doarXiv{1810.11394}

\bibitem[{{Bucciantini}(2018)}]{2018MNRAS.478.2074B}
{Bucciantini}, N. 2018, \mnras, 478, 2074, \dodoi{10.1093/mnras/sty1199}

\bibitem[{{Cao} {et~al.}(2021){Cao}, {Aharonian}, {An}, {Axikegu}, {Bai}, {Bao}, {Bastieri}, {Bi}, {Bi}, {Cai}, {Cai}, {Cao}, {Chang}, {Chang}, {Chang}, {Chen}, {Chen}, {Chen}, {Chen}, {Chen}, {Chen}, {Chen}, {Chen}, {Chen}, {Chen}, {Chen}, {Chen}, {Chen}, {Cheng}, {Cheng}, {Cui}, {Cui}, {Cui}, {Dai}, {Dai}, {Dai}, {Danzengluobu}, {della Volpe}, {D'Ettorre Piazzoli}, {Dong}, {Fan}, {Fan}, {Fan}, {Fang}, {Fang}, {Feng}, {Feng}, {Feng}, {Feng}, {Gao}, {Gao}, {Gao}, {Gao}, {Ge}, {Geng}, {Gong}, {Gou}, {Gu}, {Guo}, {Guo}, {Guo}, {Guo}, {Han}, {He}, {He}, {He}, {He}, {He}, {He}, {Heller}, {Hor}, {Hou}, {Hou}, {Hu}, {Hu}, {Hu}, {Hu}, {Huang}, {Huang}, {Huang}, {Huang}, {Huang}, {Ji}, {Ji}, {Jia}, {Jiang}, {Jiang}, {Jin}, {Kuleshov}, {Levochkin}, {Li}, {Li}, {Li}, {Li}, {Li}, {Li}, {Li}, {Li}, {Li}, {Li}, {Li}, {Li}, {Li}, {Li}, {Li}, {Li}, {Li}, {Liang}, {Liang}, {Lin}, {Liu}, {Liu}, {Liu}, {Liu}, {Liu}, {Liu}, {Liu}, {Liu}, {Liu}, {Liu}, {Liu}, {Liu}, {Liu}, {Liu}, {Liu}, {Long}, {Lu}, {Lv}, {Ma}, {Ma}, {Ma},
  {Mao}, {Masood}, {Mitthumsiri}, {Montaruli}, {Nan}, {Pang}, {Pattarakijwanich}, {Pei}, {Qi}, {Ruffolo}, {Rulev}, {S{\'a}iz}, {Shao}, {Shchegolev}, {Sheng}, {Shi}, {Song}, {Stenkin}, {Stepanov}, {Sun}, {Sun}, {Sun}, {Tam}, {Tang}, {Tian}, {Wang}, {Wang}, {Wang}, {Wang}, {Wang}, {Wang}, {Wang}, {Wang}, {Wang}, {Wang}, {Wang}, {Wang}, {Wang}, {Wang}, {Wang}, {Wang}, {Wang}, {Wang}, {Wang}, {Wang}, {Wang}, {Wei}, {Wei}, {Wei}, {Wen}, {Wu}, {Wu}, {Wu}, {Wu}, {Wu}, {Xi}, {Xia}, {Xia}, {Xiang}, {Xiao}, {Xiao}, {Xin}, {Xin}, {Xing}, {Xu}, {Xu}, {Xue}, {Yan}, \& {Yang}}]{2021Natur.594...33C}
{Cao}, Z., {Aharonian}, F.~A., {An}, Q., {et~al.} 2021, \nat, 594, 33, \dodoi{10.1038/s41586-021-03498-z}

\bibitem[{{Cao} {et~al.}(2024){Cao}, {Aharonian}, {An}, {Axikegu}, {Bai}, {Bao}, {Bastieri}, {Bi}, {Bi}, {Cai}, {Cao}, {Cao}, {Cao}, {Chang}, {Chang}, {Chen}, {Chen}, {Chen}, {Chen}, {Chen}, {Chen}, {Chen}, {Chen}, {Chen}, {Chen}, {Chen}, {Chen}, {Cheng}, {Cheng}, {Cui}, {Cui}, {Cui}, {Cui}, {Dai}, {Dai}, {Dai}, {Danzengluobu}, {Della Volpe}, {Dong}, {Duan}, {Fan}, {Fan}, {Fang}, {Fang}, {Feng}, {Feng}, {Feng}, {Feng}, {Feng}, {Gabici}, {Gao}, {Gao}, {Gao}, {Gao}, {Gao}, {Gao}, {Ge}, {Geng}, {Giacinti}, {Gong}, {Gou}, {Gu}, {Guo}, {Guo}, {Guo}, {Guo}, {Han}, {He}, {He}, {He}, {He}, {He}, {Heller}, {Hor}, {Hou}, {Hou}, {Hou}, {Hu}, {Hu}, {Hu}, {Huang}, {Huang}, {Huang}, {Huang}, {Huang}, {Huang}, {Huang}, {Ji}, {Jia}, {Jia}, {Jiang}, {Jiang}, {Jiang}, {Jin}, {Kang}, {Ke}, {Kuleshov}, {Kurinov}, {Li}, {Li}, {Li}, {Li}, {Li}, {Li}, {Li}, {Li}, {Li}, {Li}, {Li}, {Li}, {Li}, {Li}, {Li}, {Li}, {Li}, {Li}, {Li}, {Liang}, {Liang}, {Lin}, {Liu}, {Liu}, {Liu}, {Liu}, {Liu}, {Liu}, {Liu}, {Liu}, {Liu}, {Liu}, {Liu},
  {Liu}, {Liu}, {Liu}, {Lu}, {Luo}, {Lv}, {Ma}, {Ma}, {Ma}, {Mao}, {Min}, {Mitthumsiri}, {Mu}, {Nan}, {Neronov}, {Ou}, {Pang}, {Pattarakijwanich}, {Pei}, {Qi}, {Qi}, {Qiao}, {Qin}, {Ruffolo}, {S{\'a}iz}, {Semikoz}, {Shao}, {Shao}, {Shchegolev}, {Sheng}, {Shu}, {Song}, {Stenkin}, {Stepanov}, {Su}, {Sun}, {Sun}, {Sun}, {Tam}, {Tang}, {Tang}, {Tian}, {Wang}, {Wang}, {Wang}, {Wang}, {Wang}, {Wang}, {Wang}, {Wang}, {Wang}, {Wang}, {Wang}, {Wang}, {Wang}, {Wang}, {Wang}, {Wang}, {Wang}, {Wang}, {Wang}, {Wang}, {Wang}, {Wei}, {Wei}, {Wei}, {Wen}, {Wu}, \& {Wu}}]{2024ApJS..271...25C}
{Cao}, Z., {Aharonian}, F., {An}, Q., {et~al.} 2024, \apjs, 271, 25, \dodoi{10.3847/1538-4365/acfd29}

\bibitem[{{Cao} {et~al.}(2025{\natexlab{a}}){Cao}, {Aharonian}, {Bai}, {Bao}, {Bastieri}, {Bi}, {Bi}, {Bian}, {Bukevich}, {Cai}, {Cao}, {Cao}, {Chang}, {Chang}, {Chen}, {Chen}, {Chen}, {Chen}, {Chen}, {Chen}, {Chen}, {Chen}, {Chen}, {Chen}, {Chen}, {Chen}, {Chen}, {Chen}, {Chen}, {Cheng}, {Cheng}, {Chu}, {Cui}, {Cui}, {Cui}, {Dai}, {Dai}, {Dai}, {Danzengluobu}, {Diao}, {Dong}, {Duan}, {Fan}, {Fan}, {Fang}, {Fang}, {Fang}, {Feng}, {Feng}, {Feng}, {Feng}, {Feng}, {Feng}, {Gabici}, {Gao}, {Gao}, {Gao}, {Gao}, {Gao}, {Ge}, {Geng}, {Giacinti}, {Gong}, {Gou}, {Gu}, {Guo}, {Guo}, {Guo}, {Guo}, {Guo}, {Han}, {Hannuksela}, {Hasan}, {He}, {He}, {He}, {He}, {He}, {Hern{\'a}ndez-Cadena}, {Hou}, {Hou}, {Hou}, {Hu}, {Hu}, {Huang}, {Huang}, {Huang}, {Huang}, {Huang}, {Huang}, {Huang}, {Huang}, {Ji}, {Jia}, {Jia}, {Jiang}, {Jiang}, {Jiang}, {Jiang}, {Jin}, {Kaci}, {Kang}, {Karpikov}, {Khangulyan}, {Kuleshov}, {Kurinov}, {Li}, {Li}, {Li}, {Li}, {Li}, {Li}, {Li}, {Li}, {Li}, {Li}, {Li}, {Li}, {Li}, {Li}, {Li}, {Li}, {Li},
  {Li}, {Li}, {Li}, {Liu}, {Liu}, {Liu}, {Liu}, {Liu}, {Liu}, {Liu}, {Liu}, {Liu}, {Liu}, {Liu}, {Liu}, {Liu}, {Liu}, {Liu}, {Liu}, {Liu}, {Lou}, {Luo}, {Lv}, {Ma}, {Ma}, {Ma}, {Mao}, {Min}, {Mitthumsiri}, {Mou}, {Mu}, {Nan}, {Neronov}, {Ng}, {Ni}, {Nie}, {Ou}, {Pattarakijwanich}, {Pei}, {Qi}, {Qi}, {Qin}, {Raza}, {Ren}, {Ruffolo}, {S{\'a}iz}, {Saeed}, {Semikoz}, {Shao}, {Shchegolev}, {Shen}, {Sheng}, {Shi}, {Shu}, {Song}, {Stenkin}, {Stepanov}, {Su}, {Sun}, {Sun}, {Sun}, {Takata}, {Tan}, {Tang}, {Tang}, {Tang}, {Tian}, {Tong}, {Wang}, {Wang}, {Wang}, {Wang}, {Wang}, {Wang}, {Wang}, {Wang}, \& {Wang}}]{2025arXiv250215447C}
{Cao}, Z., {Aharonian}, F., {Bai}, Y.~X., {et~al.} 2025{\natexlab{a}}, arXiv e-prints, arXiv:2502.15447, \dodoi{10.48550/arXiv.2502.15447}

\bibitem[{{Cao} {et~al.}(2025{\natexlab{b}}){Cao}, {Aharonian}, {Bai}, {Bao}, {Bastieri}, {Bi}, {Bi}, {Bian WenYi}, {Butkevich}, {Cai}, {Cao}, {Cao}, {Chang}, {Chang}, {Aming Chen}, {Chen}, {Guo-Hai Chen}, {Huaxi Chen}, {Chen}, {Chen}, {Chen}, {Chen}, {Chen}, {Chen}, {Chen}, {Chen}, {Chen}, {Chen}, {Chen}, {Chen}, {Cheng}, {Cheng}, {Chung Chu}, {Cui}, {Cui}, {Cui}, {Cui}, {Dai}, {Dai}, {Dai}, {. }, {Diao}, {Dong}, {Duan}, {Fan}, {Fan}, {Fang}, {Fang}, {Fang}, {Feng}, {Feng}, {Feng}, {Feng}, {Feng}, {Feng}, {Feng}, {Gabici}, {Gao}, {Gao}, {Gao}, {Gao}, {Gao}, {Ge}, {Ge}, {Geng}, {Giacinti}, {Gong}, {Gou}, {Gu}, {Guo}, {Guo}, {Guo}, {Guo}, {Guo}, {Han}, {Hannuksela}, {Hasan}, {He}, {He}, {He}, {He}, {He}, {Hern{\'a}ndez-Cadena}, {Hou}, {Hou}, {Hou}, {Hu}, {Hu}, {Huang}, {Huang}, {Huang}, {Huang}, {Huang}, {Huang}, {Huang}, {Huang}, {Huang}, {Ji}, {Jia}, {Jia}, {Jiang}, {Jiang}, {Jiang}, {Jiang}, {Jin}, {Kaci}, {Kang}, {Karpikov}, {Khangulyan}, {Kuleshov}, {Kurinov}, {Li}, {Li}, {Cong Li}, {Li}, {Li}, {Li},
  {Li}, {Li}, {Li}, {Li}, {Li}, {Li}, {Li}, {Li}, {Li}, {Li}, {Li}, {Li}, {Li}, {Li}, {Li}, {Liang}, {Liang}, {Lin}, {Liu}, {Liu}, {Liu}, {Liu}, {Liu}, {Liu}, {Liu}, {Liu}, {Liu}, {Liu}, {Liu}, {Liu}, {Liu}, {Liu}, {Liu}, {Liu}, {Liu}, {Lou}, {Luo}, {Luo}, {Lv}, {Ma}, {Ma}, {Ma}, {Mao}, {Min}, {Mitthumsiri}, {Mou}, {Mu}, {Neronov}, {Ng}, {Ni}, {Nie}, {Lejian Ou}, {Pattarakijw anich}, {Pei}, {Qi}, {Qi}, {Qin}, {Raza}, {Ren}, {Ruffolo}, {S{\'a}iz}, {Semikoz}, {Shao}, {Shchegolev}, {Shen}, {Sheng}, {Shi}, {Shu}, {Song}, {Stepanov Stepanov}, {Su}, {Sun}, {Sun}, {Sun}, {Sun}, {Sun}, {Nabeel Tabasam}, {Takata}, {Tam}, {Tan}, {Tang}, {Tang}, \& {Tang}}]{2025arXiv251006786C}
{Cao}, Z., {Aharonian}, F., {Bai}, Y., {et~al.} 2025{\natexlab{b}}, arXiv e-prints, arXiv:2510.06786, \dodoi{10.48550/arXiv.2510.06786}

\bibitem[{{Chambers} {et~al.}(2016){Chambers}, {Magnier}, {Metcalfe}, {Flewelling}, {Huber}, {Waters}, {Denneau}, {Draper}, {Farrow}, {Finkbeiner}, {Holmberg}, {Koppenhoefer}, {Price}, {Rest}, {Saglia}, {Schlafly}, {Smartt}, {Sweeney}, {Wainscoat}, {Burgett}, {Chastel}, {Grav}, {Heasley}, {Hodapp}, {Jedicke}, {Kaiser}, {Kudritzki}, {Luppino}, {Lupton}, {Monet}, {Morgan}, {Onaka}, {Shiao}, {Stubbs}, {Tonry}, {White}, {Ba{\~n}ados}, {Bell}, {Bender}, {Bernard}, {Boegner}, {Boffi}, {Botticella}, {Calamida}, {Casertano}, {Chen}, {Chen}, {Cole}, {Deacon}, {Frenk}, {Fitzsimmons}, {Gezari}, {Gibbs}, {Goessl}, {Goggia}, {Gourgue}, {Goldman}, {Grant}, {Grebel}, {Hambly}, {Hasinger}, {Heavens}, {Heckman}, {Henderson}, {Henning}, {Holman}, {Hopp}, {Ip}, {Isani}, {Jackson}, {Keyes}, {Koekemoer}, {Kotak}, {Le}, {Liska}, {Long}, {Lucey}, {Liu}, {Martin}, {Masci}, {McLean}, {Mindel}, {Misra}, {Morganson}, {Murphy}, {Obaika}, {Narayan}, {Nieto-Santisteban}, {Norberg}, {Peacock}, {Pier}, {Postman}, {Primak}, {Rae}, {Rai},
  {Riess}, {Riffeser}, {Rix}, {R{\"o}ser}, {Russel}, {Rutz}, {Schilbach}, {Schultz}, {Scolnic}, {Strolger}, {Szalay}, {Seitz}, {Small}, {Smith}, {Soderblom}, {Taylor}, {Thomson}, {Taylor}, {Thakar}, {Thiel}, {Thilker}, {Unger}, {Urata}, {Valenti}, {Wagner}, {Walder}, {Walter}, {Watters}, {Werner}, {Wood-Vasey}, \& {Wyse}}]{2016arXiv161205560C}
{Chambers}, K.~C., {Magnier}, E.~A., {Metcalfe}, N., {et~al.} 2016, arXiv e-prints, arXiv:1612.05560.
\newblock \doarXiv{1612.05560}

\bibitem[{{Condon} {et~al.}(1998){Condon}, {Cotton}, {Greisen}, {Yin}, {Perley}, {Taylor}, \& {Broderick}}]{1998AJ....115.1693C}
{Condon}, J.~J., {Cotton}, W.~D., {Greisen}, E.~W., {et~al.} 1998, \aj, 115, 1693, \dodoi{10.1086/300337}

\bibitem[{{Cordes} \& {Lazio}(2002)}]{2002astro.ph..7156C}
{Cordes}, J.~M., \& {Lazio}, T.~J.~W. 2002, arXiv e-prints, astro, \dodoi{10.48550/arXiv.astro-ph/0207156}

\bibitem[{Croston {et~al.}(2005)Croston, Hardcastle, Harris, Belsole, Birkinshaw, \& Worrall}]{Croston2005}
Croston, J.~H., Hardcastle, M.~J., Harris, D.~E., {et~al.} 2005, The Astrophysical Journal, 626, 733, \dodoi{10.1086/430171}

\bibitem[{{Cutri} {et~al.}(2021){Cutri}, {Wright}, {Conrow}, {Fowler}, {Eisenhardt}, {Grillmair}, {Kirkpatrick}, {Masci}, {McCallon}, {Wheelock}, {Fajardo-Acosta}, {Yan}, {Benford}, {Harbut}, {Jarrett}, {Lake}, {Leisawitz}, {Ressler}, {Stanford}, {Tsai}, {Liu}, {Helou}, {Mainzer}, {Gettngs}, {Gonzalez}, {Hoffman}, {Marsh}, {Padgett}, {Skrutskie}, {Beck}, {Papin}, \& {Wittman}}]{2014yCat.2328....0C}
{Cutri}, R.~M., {Wright}, E.~L., {Conrow}, T., {et~al.} 2021, VizieR Online Data Catalog, II/328

\bibitem[{{De La Torre Luque} {et~al.}(2022){De La Torre Luque}, {Fornieri}, \& {Linden}}]{2022PhRvD.106l3033D}
{De La Torre Luque}, P., {Fornieri}, O., \& {Linden}, T. 2022, \prd, 106, 123033, \dodoi{10.1103/PhysRevD.106.123033}

\bibitem[{{de O{\~n}a Wilhelmi} {et~al.}(2022){de O{\~n}a Wilhelmi}, {L{\'o}pez-Coto}, {Amato}, \& {Aharonian}}]{2022ApJ...930L...2D}
{de O{\~n}a Wilhelmi}, E., {L{\'o}pez-Coto}, R., {Amato}, E., \& {Aharonian}, F. 2022, \apjl, 930, L2, \dodoi{10.3847/2041-8213/ac66cf}

\bibitem[{{de Vries} \& {Romani}(2022)}]{2022ApJ...928...39D}
{de Vries}, M., \& {Romani}, R.~W. 2022, \apj, 928, 39, \dodoi{10.3847/1538-4357/ac5739}

\bibitem[{{de Vries} {et~al.}(2022){de Vries}, {Romani}, {Kargaltsev}, {Pavlov}, {Posselt}, {Slane}, {Bucciantini}, {Ng}, \& {Klingler}}]{2022ApJ...939...70D}
{de Vries}, M., {Romani}, R.~W., {Kargaltsev}, O., {et~al.} 2022, \apj, 939, 70, \dodoi{10.3847/1538-4357/ac9794}

\bibitem[{{Dekker} {et~al.}(2024){Dekker}, {Holst}, {Hooper}, {Leone}, {Simon}, \& {Xiao}}]{2024PhRvD.109h3026D}
{Dekker}, A., {Holst}, I., {Hooper}, D., {et~al.} 2024, \prd, 109, 083026, \dodoi{10.1103/PhysRevD.109.083026}

\bibitem[{{Dermer} \& {Menon}(2009)}]{2009herb.book.....D}
{Dermer}, C.~D., \& {Menon}, G. 2009, {High Energy Radiation from Black Holes: Gamma Rays, Cosmic Rays, and Neutrinos}

\bibitem[{{Evans} {et~al.}(2010){Evans}, {Primini}, {Glotfelty}, {Anderson}, {Bonaventura}, {Chen}, {Davis}, {Doe}, {Evans}, {Fabbiano}, {Galle}, {Gibbs}, {Grier}, {Hain}, {Hall}, {Harbo}, {He}, {Houck}, {Karovska}, {Kashyap}, {Lauer}, {McCollough}, {McDowell}, {Miller}, {Mitschang}, {Morgan}, {Mossman}, {Nichols}, {Nowak}, {Plummer}, {Refsdal}, {Rots}, {Siemiginowska}, {Sundheim}, {Tibbetts}, {Van Stone}, {Winkelman}, \& {Zografou}}]{2010ApJS..189...37E}
{Evans}, I.~N., {Primini}, F.~A., {Glotfelty}, K.~J., {et~al.} 2010, \apjs, 189, 37, \dodoi{10.1088/0067-0049/189/1/37}

\bibitem[{{Evans} {et~al.}(2020){Evans}, {Primini}, {Miller}, {Evans}, {Allen}, {Anderson}, {Becker}, {Budynkiewicz}, {Burke}, {Chen}, {Civano}, {D'Abrusco}, {Doe}, {Fabbiano}, {Martinez Galarza}, {Gibbs}, {Glotfelty}, {Graessle}, {Grier}, {Hain}, {Hall}, {Harbo}, {Houck}, {Lauer}, {Laurino}, {Lee}, {McCollough}, {McDowell}, {McLaughlin}, {Morgan}, {Mossman}, {Nguyen}, {Nichols}, {Nowak}, {Paxson}, {Perdikeas}, {Plummer}, {Rots}, {Siemiginowska}, {Sundheim}, {Thong}, {Tibbetts}, {Van Stone}, {Winkelman}, \& {Zografou}}]{2020AAS...23515405E}
{Evans}, I.~N., {Primini}, F.~A., {Miller}, J.~B., {et~al.} 2020, in American Astronomical Society Meeting Abstracts, Vol. 235, American Astronomical Society Meeting Abstracts \#235, 154.05

\bibitem[{{Fang} {et~al.}(2019){Fang}, {Bi}, \& {Yin}}]{2019MNRAS.488.4074F}
{Fang}, K., {Bi}, X.-J., \& {Yin}, P.-F. 2019, \mnras, 488, 4074, \dodoi{10.1093/mnras/stz1974}

\bibitem[{{Gaia Collaboration} {et~al.}(2023){Gaia Collaboration}, {Vallenari}, {Brown}, {Prusti}, {de Bruijne}, {Arenou}, {Babusiaux}, {Biermann}, {Creevey}, {Ducourant}, {Evans}, {Eyer}, {Guerra}, {Hutton}, {Jordi}, {Klioner}, {Lammers}, {Lindegren}, {Luri}, {Mignard}, {Panem}, {Pourbaix}, {Randich}, {Sartoretti}, {Soubiran}, {Tanga}, {Walton}, {Bailer-Jones}, {Bastian}, {Drimmel}, {Jansen}, {Katz}, {Lattanzi}, {van Leeuwen}, {Bakker}, {Cacciari}, {Casta{\~n}eda}, {De Angeli}, {Fabricius}, {Fouesneau}, {Fr{\'e}mat}, {Galluccio}, {Guerrier}, {Heiter}, {Masana}, {Messineo}, {Mowlavi}, {Nicolas}, {Nienartowicz}, {Pailler}, {Panuzzo}, {Riclet}, {Roux}, {Seabroke}, {Sordo}, {Th{\'e}venin}, {Gracia-Abril}, {Portell}, {Teyssier}, {Altmann}, {Andrae}, {Audard}, {Bellas-Velidis}, {Benson}, {Berthier}, {Blomme}, {Burgess}, {Busonero}, {Busso}, {C{\'a}novas}, {Carry}, {Cellino}, {Cheek}, {Clementini}, {Damerdji}, {Davidson}, {de Teodoro}, {Nu{\~n}ez Campos}, {Delchambre}, {Dell'Oro}, {Esquej},
  {Fern{\'a}ndez-Hern{\'a}ndez}, {Fraile}, {Garabato}, {Garc{\'\i}a-Lario}, {Gosset}, {Haigron}, {Halbwachs}, {Hambly}, {Harrison}, {Hern{\'a}ndez}, {Hestroffer}, {Hodgkin}, {Holl}, {Jan{\ss}en}, {Jevardat de Fombelle}, {Jordan}, {Krone-Martins}, {Lanzafame}, {L{\"o}ffler}, {Marchal}, {Marrese}, {Moitinho}, {Muinonen}, {Osborne}, {Pancino}, {Pauwels}, {Recio-Blanco}, {Reyl{\'e}}, {Riello}, {Rimoldini}, {Roegiers}, {Rybizki}, {Sarro}, {Siopis}, {Smith}, {Sozzetti}, {Utrilla}, {van Leeuwen}, {Abbas}, {{\'A}brah{\'a}m}, {Abreu Aramburu}, {Aerts}, {Aguado}, {Ajaj}, {Aldea-Montero}, {Altavilla}, {{\'A}lvarez}, {Alves}, {Anders}, {Anderson}, {Anglada Varela}, {Antoja}, {Baines}, {Baker}, {Balaguer-N{\'u}{\~n}ez}, {Balbinot}, {Balog}, {Barache}, {Barbato}, {Barros}, {Barstow}, {Bartolom{\'e}}, {Bassilana}, {Bauchet}, {Becciani}, {Bellazzini}, {Berihuete}, {Bernet}, {Bertone}, {Bianchi}, {Binnenfeld}, {Blanco-Cuaresma}, {Blazere}, {Boch}, {Bombrun}, {Bossini}, {Bouquillon}, {Bragaglia}, {Bramante}, {Breedt},
  {Bressan}, {Brouillet}, {Brugaletta}, {Bucciarelli}, {Burlacu}, {Butkevich}, {Buzzi}, {Caffau}, {Cancelliere}, {Cantat-Gaudin}, {Carballo}, {Carlucci}, {Carnerero}, {Carrasco}, {Casamiquela}, {Castellani}, {Castro-Ginard}, {Chaoul}, {Charlot}, {Chemin}, {Chiaramida}, {Chiavassa}, {Chornay}, {Comoretto}, {Contursi}, {Cooper}, {Cornez}, {Cowell}, {Crifo}, {Cropper}, {Crosta}, {Crowley}, {Dafonte}, {Dapergolas}, {David}, {David}, {de Laverny}, {De Luise}, \& {De March}}]{2023A&A...674A...1G}
{Gaia Collaboration}, {Vallenari}, A., {Brown}, A.~G.~A., {et~al.} 2023, \aap, 674, A1, \dodoi{10.1051/0004-6361/202243940}

\bibitem[{{Gordon} {et~al.}(2021){Gordon}, {Boyce}, {O'Dea}, {Rudnick}, {Andernach}, {Vantyghem}, {Baum}, {Bui}, {Dionyssiou}, {Safi-Harb}, \& {Sander}}]{2021ApJS..255...30G}
{Gordon}, Y.~A., {Boyce}, M.~M., {O'Dea}, C.~P., {et~al.} 2021, \apjs, 255, 30, \dodoi{10.3847/1538-4365/ac05c0}

\bibitem[{{Gu{\'e}pin} {et~al.}(2020){Gu{\'e}pin}, {Cerutti}, \& {Kotera}}]{2020A&A...635A.138G}
{Gu{\'e}pin}, C., {Cerutti}, B., \& {Kotera}, K. 2020, \aap, 635, A138, \dodoi{10.1051/0004-6361/201936816}

\bibitem[{Hardcastle \& Croston(2020)}]{Hardcastle2020}
Hardcastle, M.~J., \& Croston, J.~H. 2020, New Astronomy Reviews, 88, 101539, \dodoi{10.1016/j.newar.2020.101539}

\bibitem[{{Kargaltsev} {et~al.}(2015){Kargaltsev}, {Cerutti}, {Lyubarsky}, \& {Striani}}]{2015SSRv..191..391K}
{Kargaltsev}, O., {Cerutti}, B., {Lyubarsky}, Y., \& {Striani}, E. 2015, \ssr, 191, 391, \dodoi{10.1007/s11214-015-0171-x}

\bibitem[{{Kargaltsev} {et~al.}(2008){Kargaltsev}, {Misanovic}, {Pavlov}, {Wong}, \& {Garmire}}]{2008ApJ...684..542K}
{Kargaltsev}, O., {Misanovic}, Z., {Pavlov}, G.~G., {Wong}, J.~A., \& {Garmire}, G.~P. 2008, \apj, 684, 542, \dodoi{10.1086/589145}

\bibitem[{{Khangulyan} {et~al.}(2014){Khangulyan}, {Aharonian}, \& {Kelner}}]{2014ApJ...783..100K}
{Khangulyan}, D., {Aharonian}, F.~A., \& {Kelner}, S.~R. 2014, \apj, 783, 100, \dodoi{10.1088/0004-637X/783/2/100}

\bibitem[{{Klingler} {et~al.}(2023){Klingler}, {Hare}, {Kargaltsev}, {Pavlov}, \& {Tomsick}}]{2023ApJ...950..177K}
{Klingler}, N., {Hare}, J., {Kargaltsev}, O., {Pavlov}, G.~G., \& {Tomsick}, J. 2023, \apj, 950, 177, \dodoi{10.3847/1538-4357/accd60}

\bibitem[{{Klingler} {et~al.}(2016){Klingler}, {Rangelov}, {Kargaltsev}, {Pavlov}, {Romani}, {Posselt}, {Slane}, {Temim}, {Ng}, {Bucciantini}, {Bykov}, {Swartz}, \& {Buehler}}]{2016ApJ...833..253K}
{Klingler}, N., {Rangelov}, B., {Kargaltsev}, O., {et~al.} 2016, \apj, 833, 253, \dodoi{10.3847/1538-4357/833/2/253}

\bibitem[{{Lacy} {et~al.}(2020){Lacy}, {Baum}, {Chandler}, {Chatterjee}, {Clarke}, {Deustua}, {English}, {Farnes}, {Gaensler}, {Gugliucci}, {Hallinan}, {Kent}, {Kimball}, {Law}, {Lazio}, {Marvil}, {Mao}, {Medlin}, {Mooley}, {Murphy}, {Myers}, {Osten}, {Richards}, {Rosolowsky}, {Rudnick}, {Schinzel}, {Sivakoff}, {Sjouwerman}, {Taylor}, {White}, {Wrobel}, {Andernach}, {Beasley}, {Berger}, {Bhatnager}, {Birkinshaw}, {Bower}, {Brandt}, {Brown}, {Burke-Spolaor}, {Butler}, {Comerford}, {Demorest}, {Fu}, {Giacintucci}, {Golap}, {G{\"u}th}, {Hales}, {Hiriart}, {Hodge}, {Horesh}, {Ivezi{\'c}}, {Jarvis}, {Kamble}, {Kassim}, {Liu}, {Loinard}, {Lyons}, {Masters}, {Mezcua}, {Moellenbrock}, {Mroczkowski}, {Nyland}, {O'Dea}, {O'Sullivan}, {Peters}, {Radford}, {Rao}, {Robnett}, {Salcido}, {Shen}, {Sobotka}, {Witz}, {Vaccari}, {van Weeren}, {Vargas}, {Williams}, \& {Yoon}}]{2020PASP..132c5001L}
{Lacy}, M., {Baum}, S.~A., {Chandler}, C.~J., {et~al.} 2020, \pasp, 132, 035001, \dodoi{10.1088/1538-3873/ab63eb}

\bibitem[{{LHAASO Collaboration} {et~al.}(2025){LHAASO Collaboration}, {Cao}, {Aharonian}, \& {An}}]{2025100802}
{LHAASO Collaboration}, {Cao}, Z., {Aharonian}, F., \& {An}, Q. 2025, The Innovation, 6, 100802, \dodoi{https://doi.org/10.1016/j.xinn.2025.100802}

\bibitem[{{Lhaaso Collaboration} {et~al.}(2021){Lhaaso Collaboration}, {Cao}, {Aharonian}, {An}, {Axikegu}, {Bai}, {Bai}, {Bao}, {Bastieri}, {Bi}, {Bi}, {Cai}, {Cai}, {Cao}, {Chang}, {Chang}, {Chen}, {Chen}, {Chen}, {Chen}, {Chen}, {Chen}, {Chen}, {Chen}, {Chen}, {Chen}, {Chen}, {Chen}, {Chen}, {Chen}, {Cheng}, {Cheng}, {Cui}, {Cui}, {Cui}, {D'Ettorre Piazzoli}, {Dai}, {Dai}, {Dai}, {Danzengluobu}, {Della Volpe}, {Dong}, {Duan}, {Fan}, {Fan}, {Fan}, {Fang}, {Fang}, {Feng}, {Feng}, {Feng}, {Feng}, {Gao}, {Gao}, {Gao}, {Gao}, {Gao}, {Ge}, {Geng}, {Gong}, {Gou}, {Gu}, {Guo}, {Guo}, {Guo}, {Guo}, {Guo}, {Han}, {He}, {He}, {He}, {He}, {He}, {He}, {Heller}, {Hor}, {Hou}, {Hou}, {Hu}, {Hu}, {Hu}, {Hu}, {Huang}, {Huang}, {Huang}, {Huang}, {Huang}, {Huang}, {Ji}, {Ji}, {Jia}, {Jiang}, {Jiang}, {Jin}, {Ke}, {Kuleshov}, {Levochkin}, {Li}, {Li}, {Li}, {Li}, {Li}, {Li}, {Li}, {Li}, {Li}, {Li}, {Li}, {Li}, {Li}, {Li}, {Li}, {Li}, {Li}, {Li}, {Liang}, {Liang}, {Lin}, {Liu}, {Liu}, {Liu}, {Liu}, {Liu}, {Liu}, {Liu}, {Liu},
  {Liu}, {Liu}, {Liu}, {Liu}, {Liu}, {Liu}, {Liu}, {Liu}, {Long}, {Lu}, {Lv}, {Ma}, {Ma}, {Ma}, {Mao}, {Masood}, {Min}, {Mitthumsiri}, {Montaruli}, {Nan}, {Pang}, {Pattarakijwanich}, {Pei}, {Qi}, {Qi}, {Qiao}, {Qin}, {Ruffolo}, {Rulev}, {Saiz}, {Shao}, {Shchegolev}, {Sheng}, {Shi}, {Song}, {Stenkin}, {Stepanov}, {Su}, {Sun}, {Sun}, {Sun}, {Tam}, {Tang}, {Tian}, {Wang}, {Wang}, {Wang}, {Wang}, {Wang}, {Wang}, {Wang}, {Wang}, {Wang}, {Wang}, {Wang}, {Wang}, {Wang}, {Wang}, {Wang}, {Wang}, {Wang}, {Wang}, {Wang}, {Wang}, {Wang}, {Wang}, {Wei}, {Wei}, {Wei}, {Wen}, {Wu}, {Wu}, {Wu}, \& {Wu}}]{2021Sci...373..425L}
{Lhaaso Collaboration}, {Cao}, Z., {Aharonian}, F., {et~al.} 2021, Science, 373, 425, \dodoi{10.1126/science.abg5137}

\bibitem[{{Li} \& {Ma}(2023)}]{2023EPJC...83..192L}
{Li}, H., \& {Ma}, B.-Q. 2023, European Physical Journal C, 83, 192, \dodoi{10.1140/epjc/s10052-023-11334-z}

\bibitem[{{Lin} {et~al.}(2024){Lin}, {Yang}, {Hare}, {Volkov}, \& {Kargaltsev}}]{2024RNAAS...8...74L}
{Lin}, Y., {Yang}, H., {Hare}, J., {Volkov}, I., \& {Kargaltsev}, O. 2024, Research Notes of the American Astronomical Society, 8, 74, \dodoi{10.3847/2515-5172/ad324a}

\bibitem[{{Manchester} {et~al.}(2005){Manchester}, {Hobbs}, {Teoh}, \& {Hobbs}}]{2005AJ....129.1993M}
{Manchester}, R.~N., {Hobbs}, G.~B., {Teoh}, A., \& {Hobbs}, M. 2005, \aj, 129, 1993, \dodoi{10.1086/428488}

\bibitem[{{Mao} {et~al.}(2016){Mao}, {Urry}, {Massaro}, {Paggi}, {Cauteruccio}, \& {K{\"u}nzel}}]{2016ApJS..224...26M}
{Mao}, P., {Urry}, C.~M., {Massaro}, F., {et~al.} 2016, \apjs, 224, 26, \dodoi{10.3847/0067-0049/224/2/26}

\bibitem[{{McConnell} {et~al.}(2020){McConnell}, {Hale}, {Lenc}, {Banfield}, {Heald}, {Hotan}, {Leung}, {Moss}, {Murphy}, {O'Brien}, {Pritchard}, {Raja}, {Sadler}, {Stewart}, {Thomson}, {Whiting}, {Allison}, {Amy}, {Anderson}, {Ball}, {Bannister}, {Bell}, {Bock}, {Bolton}, {Bunton}, {Chippendale}, {Collier}, {Cooray}, {Cornwell}, {Diamond}, {Edwards}, {Gupta}, {Hayman}, {Heywood}, {Jackson}, {Koribalski}, {Lee-Waddell}, {McClure-Griffiths}, {Ng}, {Norris}, {Phillips}, {Reynolds}, {Roxby}, {Schinckel}, {Shields}, {Tremblay}, {Tzioumis}, {Voronkov}, \& {Westmeier}}]{2020PASA...37...48M}
{McConnell}, D., {Hale}, C.~L., {Lenc}, E., {et~al.} 2020, \pasa, 37, e048, \dodoi{10.1017/pasa.2020.41}

\bibitem[{{McGowan} {et~al.}(2006){McGowan}, {Vestrand}, {Kennea}, {Zane}, {Cropper}, \& {C{\'o}rdova}}]{2006ApJ...647.1300M}
{McGowan}, K.~E., {Vestrand}, W.~T., {Kennea}, J.~A., {et~al.} 2006, \apj, 647, 1300, \dodoi{10.1086/505522}

\bibitem[{{McLaughlin} {et~al.}(2002){McLaughlin}, {Arzoumanian}, {Cordes}, {Backer}, {Lommen}, {Lorimer}, \& {Zepka}}]{2002ApJ...564..333M}
{McLaughlin}, M.~A., {Arzoumanian}, Z., {Cordes}, J.~M., {et~al.} 2002, \apj, 564, 333, \dodoi{10.1086/324151}

\bibitem[{{Moderski} {et~al.}(2005){Moderski}, {Sikora}, {Coppi}, \& {Aharonian}}]{2005MNRAS.363..954M}
{Moderski}, R., {Sikora}, M., {Coppi}, P.~S., \& {Aharonian}, F. 2005, \mnras, 363, 954, \dodoi{10.1111/j.1365-2966.2005.09494.x}

\bibitem[{{Moran} {et~al.}(2013){Moran}, {Mignani}, {Collins}, {de Luca}, {Rea}, \& {Shearer}}]{2013MNRAS.436..401M}
{Moran}, P., {Mignani}, R.~P., {Collins}, S., {et~al.} 2013, \mnras, 436, 401, \dodoi{10.1093/mnras/stt1573}

\bibitem[{{Olmi}(2023)}]{2023Univ....9..402O}
{Olmi}, B. 2023, Universe, 9, 402, \dodoi{10.3390/universe9090402}

\bibitem[{{Olmi} {et~al.}(2024){Olmi}, {Amato}, {Bandiera}, \& {Blasi}}]{2024A&A...684L...1O}
{Olmi}, B., {Amato}, E., {Bandiera}, R., \& {Blasi}, P. 2024, \aap, 684, L1, \dodoi{10.1051/0004-6361/202449382}

\bibitem[{{Posselt} {et~al.}(2012){Posselt}, {Arumugasamy}, {Pavlov}, {Manchester}, {Shannon}, \& {Kargaltsev}}]{2012ApJ...761..117P}
{Posselt}, B., {Arumugasamy}, P., {Pavlov}, G.~G., {et~al.} 2012, \apj, 761, 117, \dodoi{10.1088/0004-637X/761/2/117}

\bibitem[{{Reynolds}(2009)}]{2009ApJ...703..662R}
{Reynolds}, S.~P. 2009, \apj, 703, 662, \dodoi{10.1088/0004-637X/703/1/662}

\bibitem[{{Reynolds} {et~al.}(2017){Reynolds}, {Pavlov}, {Kargaltsev}, {Klingler}, {Renaud}, \& {Mereghetti}}]{2017SSRv..207..175R}
{Reynolds}, S.~P., {Pavlov}, G.~G., {Kargaltsev}, O., {et~al.} 2017, \ssr, 207, 175, \dodoi{10.1007/s11214-017-0356-6}

\bibitem[{{Rigoselli} {et~al.}(2022){Rigoselli}, {Mereghetti}, {Anzuinelli}, {Keith}, {Taverna}, {Turolla}, \& {Zane}}]{2022MNRAS.513.3113R}
{Rigoselli}, M., {Mereghetti}, S., {Anzuinelli}, S., {et~al.} 2022, \mnras, 513, 3113, \dodoi{10.1093/mnras/stac1130}

\bibitem[{{Salvato} {et~al.}(2018){Salvato}, {Buchner}, {Budav{\'a}ri}, {Dwelly}, {Merloni}, {Brusa}, {Rau}, {Fotopoulou}, \& {Nandra}}]{2018MNRAS.473.4937S}
{Salvato}, M., {Buchner}, J., {Budav{\'a}ri}, T., {et~al.} 2018, \mnras, 473, 4937, \dodoi{10.1093/mnras/stx2651}

\bibitem[{{Skrutskie} {et~al.}(2006){Skrutskie}, {Cutri}, {Stiening}, {Weinberg}, {Schneider}, {Carpenter}, {Beichman}, {Capps}, {Chester}, {Elias}, {Huchra}, {Liebert}, {Lonsdale}, {Monet}, {Price}, {Seitzer}, {Jarrett}, {Kirkpatrick}, {Gizis}, {Howard}, {Evans}, {Fowler}, {Fullmer}, {Hurt}, {Light}, {Kopan}, {Marsh}, {McCallon}, {Tam}, {Van Dyk}, \& {Wheelock}}]{2006AJ....131.1163S}
{Skrutskie}, M.~F., {Cutri}, R.~M., {Stiening}, R., {et~al.} 2006, \aj, 131, 1163, \dodoi{10.1086/498708}

\bibitem[{{Smith} {et~al.}(2023){Smith}, {Abdollahi}, {Ajello}, {Bailes}, {Baldini}, {Ballet}, {Baring}, {Bassa}, {Gonzalez}, {Bellazzini}, {Berretta}, {Bhattacharyya}, {Bissaldi}, {Bonino}, {Bottacini}, {Bregeon}, {Bruel}, {Burgay}, {Burnett}, {Cameron}, {Camilo}, {Caputo}, {Caraveo}, {Cavazzuti}, {Chiaro}, {Ciprini}, {Clark}, {Cognard}, {Corongiu}, {Orestano}, {Crnogorcevic}, {Cuoco}, {Cutini}, {D'Ammando}, {de Angelis}, {DeCesar}, {De Gaetano}, {de Menezes}, {Deneva}, {de Palma}, {Di Lalla}, {Dirirsa}, {Di Venere}, {Dom{\'\i}nguez}, {Dumora}, {Fegan}, {Ferrara}, {Fiori}, {Fleischhack}, {Flynn}, {Franckowiak}, {Freire}, {Fukazawa}, {Fusco}, {Galanti}, {Gammaldi}, {Gargano}, {Gasparrini}, {Giacchino}, {Giglietto}, {Giordano}, {Giroletti}, {Green}, {Grenier}, {Guillemot}, {Guiriec}, {Gustafsson}, {Harding}, {Hays}, {Hewitt}, {Horan}, {Hou}, {Jankowski}, {Johnson}, {Johnson}, {Johnston}, {Kataoka}, {Keith}, {Kerr}, {Kramer}, {Kuss}, {Latronico}, {Lee}, {Li}, {Li}, {Limyansky}, {Longo}, {Loparco}, {Lorusso},
  {Lovellette}, {Lower}, {Lubrano}, {Lyne}, {Maan}, {Maldera}, {Manchester}, {Manfreda}, {Marelli}, {Mart{\'\i}-Devesa}, {Mazziotta}, {McEnery}, {Mereu}, {Michelson}, {Mickaliger}, {Mitthumsiri}, {Mizuno}, {Moiseev}, {Monzani}, {Morselli}, {Negro}, {Nemmen}, {Nieder}, {Nuss}, {Omodei}, {Orienti}, {Orlando}, {Ormes}, {Palatiello}, {Paneque}, {Panzarini}, {Parthasarathy}, {Persic}, {Pesce-Rollins}, {Pillera}, {Poon}, {Porter}, {Possenti}, {Principe}, {Rain{\`o}}, {Rando}, {Ransom}, {Ray}, {Razzano}, {Razzaque}, {Reimer}, {Reimer}, {Renault-Tinacci}, {Romani}, {S{\'a}nchez-Conde}, {Parkinson}, {Scotton}, {Serini}, {Sgr{\`o}}, {Shannon}, {Sharma}, {Shen}, {Siskind}, {Spandre}, {Spinelli}, {Stappers}, {Stephens}, {Suson}, {Tabassum}, {Tajima}, {Tak}, {Theureau}, {Thompson}, {Tibolla}, {Torres}, {Valverde}, {Venter}, {Wadiasingh}, {Wang}, {Wang}, {Wang}, {Weltevrede}, {Wood}, {Yan}, {Zaharijas}, {Zhang}, \& {Zhu}}]{2023ApJ...958..191S}
{Smith}, D.~A., {Abdollahi}, S., {Ajello}, M., {et~al.} 2023, \apj, 958, 191, \dodoi{10.3847/1538-4357/acee67}

\bibitem[{{Stawarz} {et~al.}(2010){Stawarz}, {Petrosian}, \& {Blandford}}]{2010ApJ...710..236S}
{Stawarz}, {\L}., {Petrosian}, V., \& {Blandford}, R.~D. 2010, \apj, 710, 236, \dodoi{10.1088/0004-637X/710/1/236}

\bibitem[{{Strader} {et~al.}(2019){Strader}, {Swihart}, {Chomiuk}, {Bahramian}, {Britt}, {Cheung}, {Dage}, {Halpern}, {Li}, {Mignani}, {Orosz}, {Peacock}, {Salinas}, {Shishkovsky}, \& {Tremou}}]{2019ApJ...872...42S}
{Strader}, J., {Swihart}, S., {Chomiuk}, L., {et~al.} 2019, \apj, 872, 42, \dodoi{10.3847/1538-4357/aafbaa}

\bibitem[{{Traulsen} {et~al.}(2020){Traulsen}, {Schwope}, {Lamer}, {Ballet}, {Carrera}, {Ceballos}, {Coriat}, {Freyberg}, {Koliopanos}, {Kurpas}, {Michel}, {Motch}, {Page}, {Watson}, \& {Webb}}]{2020A&A...641A.137T}
{Traulsen}, I., {Schwope}, A.~D., {Lamer}, G., {et~al.} 2020, \aap, 641, A137, \dodoi{10.1051/0004-6361/202037706}

\bibitem[{{Verner} {et~al.}(1996){Verner}, {Ferland}, {Korista}, \& {Yakovlev}}]{1996ApJ...465..487V}
{Verner}, D.~A., {Ferland}, G.~J., {Korista}, K.~T., \& {Yakovlev}, D.~G. 1996, \apj, 465, 487, \dodoi{10.1086/177435}

\bibitem[{{Webb} {et~al.}(2020){Webb}, {Coriat}, {Traulsen}, {Ballet}, {Motch}, {Carrera}, {Koliopanos}, {Authier}, {de la Calle}, {Ceballos}, {Colomo}, {Chuard}, {Freyberg}, {Garcia}, {Kolehmainen}, {Lamer}, {Lin}, {Maggi}, {Michel}, {Page}, {Page}, {Perea-Calderon}, {Pineau}, {Rodriguez}, {Rosen}, {Santos Lleo}, {Saxton}, {Schwope}, {Tom{\'a}s}, {Watson}, \& {Zakardjian}}]{2020A&A...641A.136W}
{Webb}, N.~A., {Coriat}, M., {Traulsen}, I., {et~al.} 2020, \aap, 641, A136, \dodoi{10.1051/0004-6361/201937353}

\bibitem[{{Wilms} {et~al.}(2000){Wilms}, {Allen}, \& {McCray}}]{2000ApJ...542..914W}
{Wilms}, J., {Allen}, A., \& {McCray}, R. 2000, \apj, 542, 914, \dodoi{10.1086/317016}

\bibitem[{Wood {et~al.}(2017)Wood, Caputo, Charles, Di~Mauro, Magill, \& Perkins}]{2017ICRC...35..824W}
Wood, M., Caputo, R., Charles, E., {et~al.} 2017, PoS, ICRC2017, 824, \dodoi{10.22323/1.301.0824}

\bibitem[{{Yan} {et~al.}(2025){Yan}, {Wu}, \& {Liu}}]{2025ApJ...987...19Y}
{Yan}, K., {Wu}, S., \& {Liu}, R.-Y. 2025, \apj, 987, 19, \dodoi{10.3847/1538-4357/add6a4}

\bibitem[{{Yang} {et~al.}(2024){Yang}, {Hare}, \& {Kargaltsev}}]{2024ApJ...971..180Y}
{Yang}, H., {Hare}, J., \& {Kargaltsev}, O. 2024, \apj, 971, 180, \dodoi{10.3847/1538-4357/ad543e}

\bibitem[{{Yang} {et~al.}(2022){Yang}, {Hare}, {Kargaltsev}, {Volkov}, {Chen}, \& {Rangelov}}]{2022ApJ...941..104Y}
{Yang}, H., {Hare}, J., {Kargaltsev}, O., {et~al.} 2022, \apj, 941, 104, \dodoi{10.3847/1538-4357/ac952b}

\bibitem[{{Yao} {et~al.}(2017){Yao}, {Manchester}, \& {Wang}}]{2017ApJ...835...29Y}
{Yao}, J.~M., {Manchester}, R.~N., \& {Wang}, N. 2017, \apj, 835, 29, \dodoi{10.3847/1538-4357/835/1/29}

\bibitem[{{Yuan} {et~al.}(2017){Yuan}, {Lin}, {Fang}, \& {Bi}}]{2017PhRvD..95h3007Y}
{Yuan}, Q., {Lin}, S.-J., {Fang}, K., \& {Bi}, X.-J. 2017, \prd, 95, 083007, \dodoi{10.1103/PhysRevD.95.083007}

\bibitem[{{Zabalza}(2015)}]{2015ICRC...34..922Z}
{Zabalza}, V. 2015, in International Cosmic Ray Conference, Vol.~34, 34th International Cosmic Ray Conference (ICRC2015), 922, \dodoi{10.22323/1.236.0922}

\bibitem[{{Zyuzin} {et~al.}(2018){Zyuzin}, {Karpova}, \& {Shibanov}}]{2018MNRAS.476.2177Z}
{Zyuzin}, D.~A., {Karpova}, A.~V., \& {Shibanov}, Y.~A. 2018, \mnras, 476, 2177, \dodoi{10.1093/mnras/sty359}

\end{thebibliography}

\appendix
\section{Machine Learning Classification with MUWCLASS}
\label{ML_appendix}
To 
identify the nature of all 
X-ray sources within the PU of 1LHAASO J1740+0948u, we 
used the MUWCLASS 
pipeline  from \citealt{2022ApJ...941..104Y, 2024ApJ...971..180Y}.
MUWCLASS 
crossmatches (see Table \ref{tab:MW_CTP}) X-ray sources to MW counterparts from Gaia DR3, Gaia EDR3 Distances, Two Micron All Sky Survey (2MASS), ALLWISE, and CatWISE catalogs using a probabilistic crossmatching method, NWAY \citep{2018MNRAS.473.4937S}, and then performs supervised machine learning classification using 
Random Forest algorithm.
The algorithm
learns from the 4XMM-based training dataset (TD), published in \cite{2024RNAAS...8...74L} or from CSC-based TD from \cite{2024ApJ...971..180Y}. 
The 4XMM-based TD  consists of 9 classes: active galactic nuclei (AGN), high-mass stars (HM-STAR)\footnote{Includes Wolf-Rayet, O, B stars.}, low-mass stars (LM-STAR), cataclysmic variables (CV), high-mass X-ray binaries (HMXB), low-mass X-ray binaries (LMXB), pulsars and isolated neutron stars (NS), non-accreting X-ray binaries (NS BIN)\footnote{Non-accreting binaries, including wide-orbit binaries with millisecond pulsars, red-back, and black widow systems \citep{2019ApJ...872...42S}.} and young stellar objects (YSO).
For CSC-based training dataset, the NS BIN class is merged with the LMXB class due to the low number of sources. MUWCLASS treats the imbalance between different classes in the TD by creating synthetic sources. 

We perform 
classifications 
for all potential counterparts as well as  a no counterpart case. For each X-ray source, we then  combine the classification probabilities  weighted by the corresponding association outcome

\begin{equation}
p_j=
    \begin{cases}
    1-p\_{\rm any}~{\rm,~no~counterpart~case}\\
    p\_{\rm any}\times p\_{\rm i}~{\rm ,}~i{\rm th~counterpart~case}
    \end{cases}
\end{equation}

where, $p$\_any is the probability of having a true counterpart among all counterparts considered, 
and $p\_i$ is the probability of the $i$th 
counterpart being the true one, reported by NWAY.
Specifically, given a class vector $\Vec{V}_{class}=[AGN, NS, ...]$, classification probability vector $\Vec{P}_{class}=[0.1, 0.34, ...]$, and classification probability dispersion\footnote{see \cite{2022ApJ...941..104Y} for details.} vector $e\_\Vec{P}_{class}=[0.03, 0.07, ...]$, the combined  classification probability vector 

\begin{equation}
    \Vec{P}=\sum_j \Vec{P}_{class, j}\times p_j
\end{equation}

and the combined classification probability dispersion vector

\begin{equation}
e\_\Vec{P}=(\sum_j (e\_\Vec{P}_{class, j}\times p_j)^2)^{1/2}.
\end{equation}

\begin{figure}
    \centering
\includegraphics[width=1.0\linewidth]{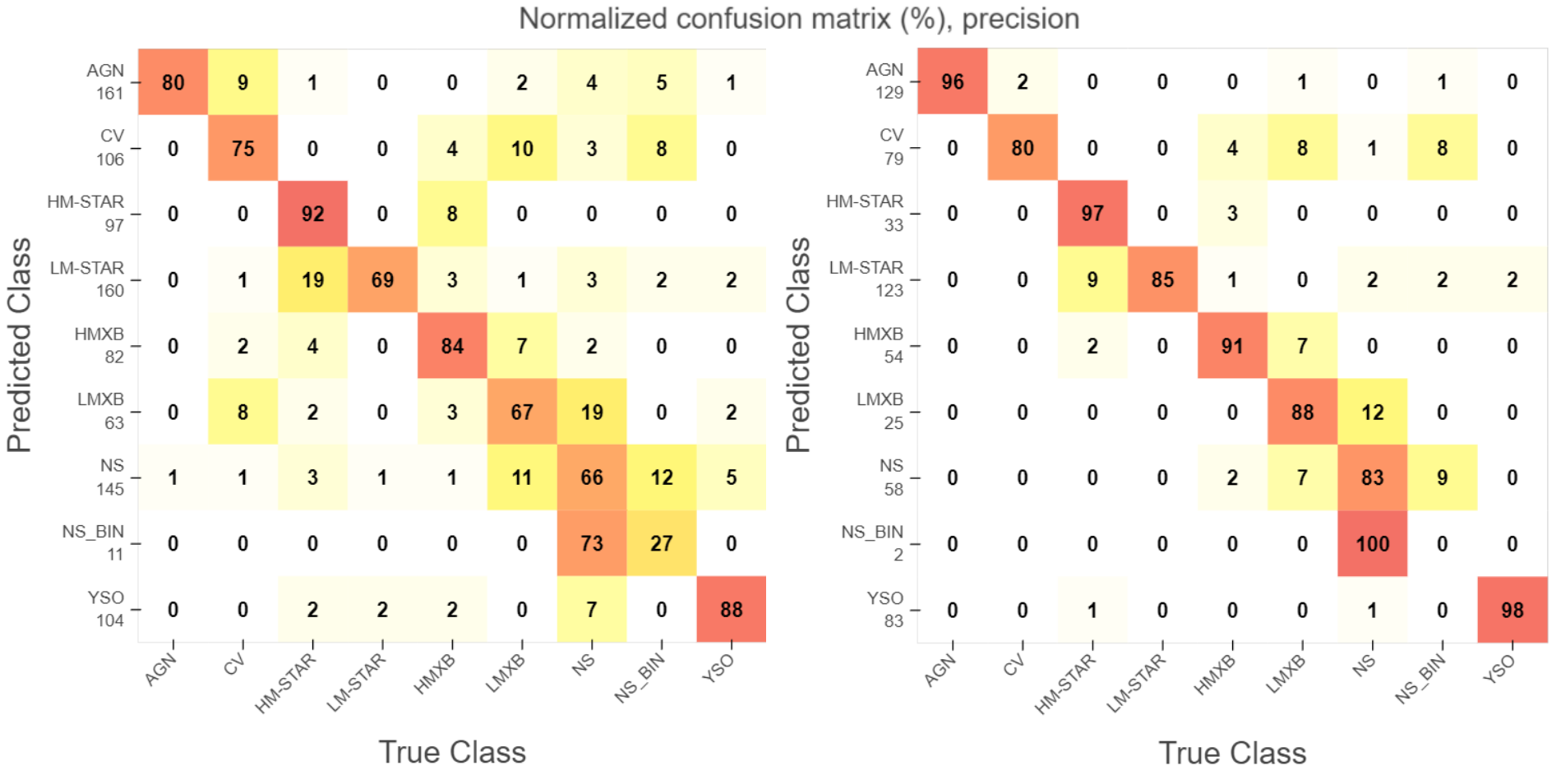}
    \caption{Normalized (over the ``'true' outcomes, i.e.,  rows) precision (the ratio of true positives to the sum of true positives and false positives) confusion matrix of the 4XMM-DR13 training dataset with any CT(left) and  CT$>$2  (right). 
    A more diagonal confusion matrix indicates a better performance. The number under each class label indicates the number of  X-ray sources from the TD  classified as that class.}
    \label{fig:CM}
\end{figure}

The overall classification probability and its dispersion, $P$ and $e\_P$ reported in Table \ref{tab:combine_table}, are the one assigned to the most likely class in $\Vec{V}_{class}$.  
The confidence threshold (CT), 
defined as the difference of the probabilities of the most likely class and the second most likely class, divided by their combined classification probability dispersion, 
is used to evaluate the reliability of classifications.
 A high CT indicates the source has high classification probability and low dispersion (see eq. 7 in \citealt{2022ApJ...941..104Y}). 
Below, we also include the confusion matrix for the {\sl XMM-Newton} training dataset (see Figure \ref{fig:CM}) from leave-one-out cross-validation, which characterizes MUWCLASS performance by class. 

\begin{table}[b!]
\vspace{-0.001cm}
 {Table A1: Multiwavelength counterparts of X-ray sources. No 2MASS counterparts was found for any of the 12 X-ray sources.} 
\vspace{-0.001cm}
\setlength{\tabcolsep}{0.04in} \footnotesize{
\begin{center}
\begin{tabular}{lccccccccc}
\hline\hline
      & name & MW\_sep & p\_i & p\_any & GAIA\_DR3Name & CATWISE\_Name & ALLWISE\_Name \\ \hline
        1&4XMM J174025.5+094703-1 & 0.97 & 1.0 & 0.935 & ~ & J174025.47+094703.1 & J174025.47+094703.7 \\ \hline
        2&2CXO J174011.3+095122-1 & 1.46 & 0.53 & 0.973 & Gaia DR3 4489358546610540416 & J174011.33+095122.6 & J174011.28+095123.1 \\ \hline
        2&2CXO J174011.3+095122-2 & 2.71 & 0.47 & 0.973 & Gaia DR3 4489358550907052672 & ~ & ~ \\ \hline
        3&4XMM J174008.4+095258-1 & 1.15 & 1.0 & 0.927 & ~ & J174008.49+095259.2 & J174008.49+095259.0 \\ \hline
        3&2CXO J174008.4+095259-1 & 0.48 & 1.0 & 0.998 & ~ & J174008.49+095259.2 & J174008.49+095259.0 \\ \hline
        4&4XMM J174001.8+094849-1 & 0.60 & 1.0 & 0.974 & ~ & J174001.91+094849.7 & ~ \\ \hline
        5&4XMM J173959.2+094814-1 & 1.16 & 1.0 & 0.942 & ~ & J173959.23+094816.0 & ~ \\ \hline
        5& 2CXO J173959.1+094817-1 & 1.83 & 0.51 & 0.979 & ~ & J173959.23+094816.0 & ~ \\ \hline
        5&2CXO J173959.1+094817-2 & 2.05 & 0.49 & 0.979 & ~ & ~ & J173959.05+094815.9 \\ \hline
        6& 4XMM J174002.3+094721-1 & 0.21 & 1.0 & 0.983 & ~ & J174002.37+094720.9 & J174002.34+094721.2 \\ \hline
        6&2CXO J174002.1+094721-1 & 3.17 & 1.0 & 0.968 & ~ & J174002.37+094720.9 & J174002.34+094721.2 \\ \hline \hline
        7c&2CXO J173952.1+094733-1 & 3.75 & 0.38 & 0.963 & ~ & J173952.37+094734.6 & J173952.42+094734.3 \\ \hline
        7c&2CXO J173952.1+094733-2 & 3.93 & 0.35 & 0.963 & ~ & J173951.92+094731.0 & ~ \\ \hline
        7c&2CXO J173952.1+094733-3 & 4.74 & 0.26 & 0.963 & Gaia DR3 4489356381946843520 & ~ & ~ \\ \hline
        8c&2CXO J173957.1+094521-1 & 2.93 & 1.0 & 0.924 & ~ & J173957.36+094521.5 & ~ \\ \hline
        
\end{tabular}
\end{center}
} \vspace{-0.2cm}
 \footnotesize
 { 
 }
 \label{tab:MW_CTP}
 \vspace{-0.0cm}
\end{table}

\end{document}